# RF-Budgeted Frame Compilation for Frequency-Multiplexed Superconducting-Qubit Control Using Qubit-Control Identity Records and a Circuit-Informed RFSoC Model

**Mingqi Ge[1], Anne-Marie Valente-Feliciano[1]**

[1]Thomas Jefferson National Accelerator Facility, Newport News, Virginia, USA, 23606

Corresponding author: Mingqi Ge (email: mingqi@jlab.org).

**ABSTRACT** Frequency-multiplexed superconducting-qubit control requires more than carrier assignment: the RF budget of a shared source can perturb multi-qubit rotations through finite bandwidth, crest factor, clipping, quantization, jitter, spurs, compression, crosstalk, and leakage. We present an RF-budgeted frame-compilation and validation workflow that combines qubit-control identity (QID) records, a MATLAB/Simulink-based circuit-informed RFSoC source-chain model, QuTiP qutrit dynamics, and Qiskit-derived algorithm workloads. QID records encode qubit-specific computational and leakage transition frequencies, pulse parameters, and drive-scale calibration, while the RF-chain profile and effective crosstalk-coupling matrix are provided as separate compiler inputs. Candidate multitone RF frames are scheduled under RF-budget constraints, propagated through the RFSoC model, decoded into computational and leakage transition frames, and evaluated in QuTiP for rotation error, leakage-aware fidelity, computational-subspace survival, and transient leakage. The studies progress from single-qutrit pulse closure to pairwise coexistence, multitone RF-frame capacity, and Bernstein-Vazirani (BV) and QAOA microwave layers extracted from Qiskit circuits. The simulations show that longer pulses improve per-frame aggregation but do not necessarily minimize time-normalized layer cost; clustered frequency maps, larger rotations, and multitone leakage stacking tighten closure. Under the nominal RF budget, a Qiskit-derived 12-qubit BV $-Y90$ layer closes in three validated four-tone frames at 240 ns, while QAOA mixer partitions vary with rotation angle and pulse duration. All reported results are model-based, decoherence-free simulation diagnostics rather than measured hardware fidelities or wiring-reduction claims.

**INDEX TERMS** superconducting qubits; microwave control; frequency-multiplexed control; RF-budgeted compilation; qubit-control identity records; RFSoC; mixed-signal modeling; circuit-informed simulation workflow; leakage-aware scheduling; crosstalk-aware control

## I. INTRODUCTION

Superconducting quantum processors are moving from isolated high-fidelity gate demonstrations toward repeated, highly parallel control cycles required for quantum error correction and logical-qubit operation [1], [2]. In this regime, microwave control becomes a system-level compilation problem. Simultaneous gates can suffer excess error from crosstalk, spectator participation, hybridization, leakage, and frequency crowding [3]. At the same time, wiring density, cryogenic heat load, package routing, and room-temperature electronics motivate shared-line, multiplexed, and RFSoC-based control architectures.

Shared-line and frequency-multiplexed microwave control have recently been explored as routes to reducing the control-resource burden in superconducting-qubit systems. Prior work includes shared microwave-drive schemes, row-column addressing, cryogenic microwave-multiplexed controllers, selective-excitation pulses on a shared control line, OFDM-inspired simultaneous-gate analysis, and multitone phase-selection methods for reducing peak microwave amplitude [4]-[9]. In parallel, RFSoC- and FPGA-based platforms such as QICK, QubiC, QubiC 2.0, and QiCells/QiController have demonstrated scalable waveform generation, readout, synchronization, feedback, programmable pulse execution, and modular firmware support for superconducting-qubit experiments [10]-[14]. These works provide important architectural, pulse-level, and hardware foundations. However, they do not by themselves answer the compiler-level admission question addressed in this paper: for a requested microwave gate layer, which tones can safely share one RF frame under explicit RF-source, DSP, leakage, crosstalk, and gate-quality budgets?

This admission problem is difficult because frequency-selective addressing in weakly anharmonic superconducting qubits must account for both computational and noncomputational transitions. A pulse intended for the computational $|0\rangle \leftrightarrow |1\rangle$ transition can also overlap the $|1\rangle \leftrightarrow |2\rangle$ leakage transition of the same qubit or nearby spectator transitions, especially for short gates and crowded frequency maps [15]-[17]. DRAG, Wah-Wah, and analytic simultaneous-gate studies showed that leakage-aware pulse design is essential in frequency-crowded multilevel systems [18], [19]. In shared-RF operation, the same physics appears at the frame level: a tone that is safe in isolation, or even in a pairwise test, can become unsafe when several tones project coherently onto a common spectator transition.

The RF source chain adds a separate class of constraints. Multitone microwave frames are affected by crest factor, full-scale headroom, clipping, finite DAC resolution, ENOB-equivalent noise, RFDC and NCO quantization, sample-clock jitter, DAC and NCO spurs, bandwidth roll-off, group delay, compression, board/package transfer response, and, for analog-IQ paths, I/Q imbalance and LO leakage [20]–[23]. Prior control-electronics studies and behavioral co-simulations have connected classical electronics specifications to qubit fidelity requirements [24], [25]. The present work uses this type of source-chain modeling in a different role: the RFSoC-informed model is embedded inside a compiler workflow that accepts, repartitions, or rejects multitone RF-frame descriptors before execution.

To support this workflow, we introduce qubit-control identity (QID) records. A QID is a compiler-side record associated with a physical qubit and contains its computational and first leakage transition frequencies, available pulse parameters, and drive-scale calibration. Crosstalk coupling and RF-chain/path-response information are supplied separately through the effective crosstalk matrix and RF-chain profile. The compiler uses these inputs jointly to determine whether requested microwave rotations can safely share an RF frame and, if not, whether the layer can be repartitioned or is hardware-limited under the configured model. The QID is used only by the compiler and is not physically encoded in or decoded by the qubit.

This record-based view is related to existing simulation, pulse-level, and calibration interfaces but serves a different role. QuTiP provides a mature framework for open-system and waveform-driven quantum-dynamics simulation [26]-[28]. Qiskit Pulse exposes pulse schedules and backend calibration information [29], and the broader Qiskit framework supports circuit construction, transformation, and algorithm-level workflow generation [30]. Software crosstalk mitigation uses characterized crosstalk to serialize unsafe operations [31], while cross-layer compiler work combines program-, gate-, and pulse-level information to reduce coherent noise [32]. QID records instead organize qubit-specific transition, pulse, and drive-calibration information as structured inputs to an RF-budgeted shared-microwave frame compiler, while calibrated crosstalk and RF-path information are supplied through separate compiler inputs.

We formulate frequency-multiplexed superconducting-qubit microwave control as an RF-budgeted frame-compilation problem. Given a requested layer of single-qubit microwave rotations, the compiler operates on a QID-aware input set comprising the per-qubit QID database, the requested gate layer, the configured RF-chain profile, and the effective crosstalk-coupling matrix. These inputs are used jointly to construct and schedule candidate multitone RF-frame descriptors. The aggregate waveform for each candidate frame is propagated through a circuit-informed direct-RF RFSoC source-chain model, decoded into computational and leakage transition frames, and evaluated using local qutrit-patch QuTiP simulations. The workflow reports RF diagnostics, Hamiltonian-level decoding diagnostics, leakage-aware average gate fidelity, a polar-unitary coherent-fidelity diagnostic, computational-subspace survival, and transient leakage. These quantities are simulation diagnostics rather than measured hardware fidelities.

The benchmark sequence progresses from single-qutrit pulse closure to pairwise same-frame coexistence, multitone RF-frame capacity, and algorithm-derived microwave layers extracted from Bernstein-Vazirani (BV) and QAOA circuits generated with Qiskit [30], [33], [34]. The main contribution is therefore not the existence of shared-line frequency-multiplexed control itself. Rather, it is an RF-budgeted frame-compilation workflow that combines qubit-specific QID records with separate RF-chain and crosstalk inputs to determine when shared-RF multitone control is admissible under both source-chain and leakage-aware qubit-dynamics constraints. Under the nominal RF budget used in this simulation study, complete 12-qubit X90 layers on the uniform, jittered, and heavy-tail maps require five validated RF frames at 120 ns and three at 240 ns, whereas the clustered map requires nine and five frames, respectively. The time-normalized aggregation is highest at 120 ns, reaching 20 qubits/GHz/μs; therefore, the pulse duration that admits the most tones per frame does not necessarily minimize the validated layer time. For the Qiskit-derived workloads, the 12-qubit BV physical −Y90 layer closes in three validated four-tone frames at 240 ns. At the same duration, QAOA mixer layers require 3, 5, 6, and 5 validated frames for physical rotation angles of 90°, 120°, 150°, and 180°, respectively. The 180° point does not enforce the rotation-axis phase criterion because azimuth is undefined at the polar target and should not be interpreted as evidence of monotonic improvement at larger angles. These results are model-based validated-partition and gate-quality diagnostics under the stated QID maps and RF profiles, not measured RFSoC hardware performance, implemented wiring reduction, or experimental qubit fidelity.

## II. QID INPUTS AND COMPILER PROBLEM DEFINITION

The compiler operates on a requested layer of single-qubit microwave rotations, a qubit-control identity (QID) database, a configured RF-chain profile, and an effective crosstalk-coupling matrix. Its output is a finite sequence of shared RF frames together with the waveform descriptors required to execute each frame. This section defines those inputs and outputs before introducing waveform encoding and physical decoding.

Figure 1 summarizes the hardware-software boundary used by the workflow. QID construction, scheduling, RF-budget evaluation, and validation-driven recompilation are host-side compiler functions. The RFSoC-inspired runtime layer executes compiled waveform descriptors and provides the source-chain waveform model used by the decoder. The plant layer is represented in this study by local qutrit-patch QuTiP validation.

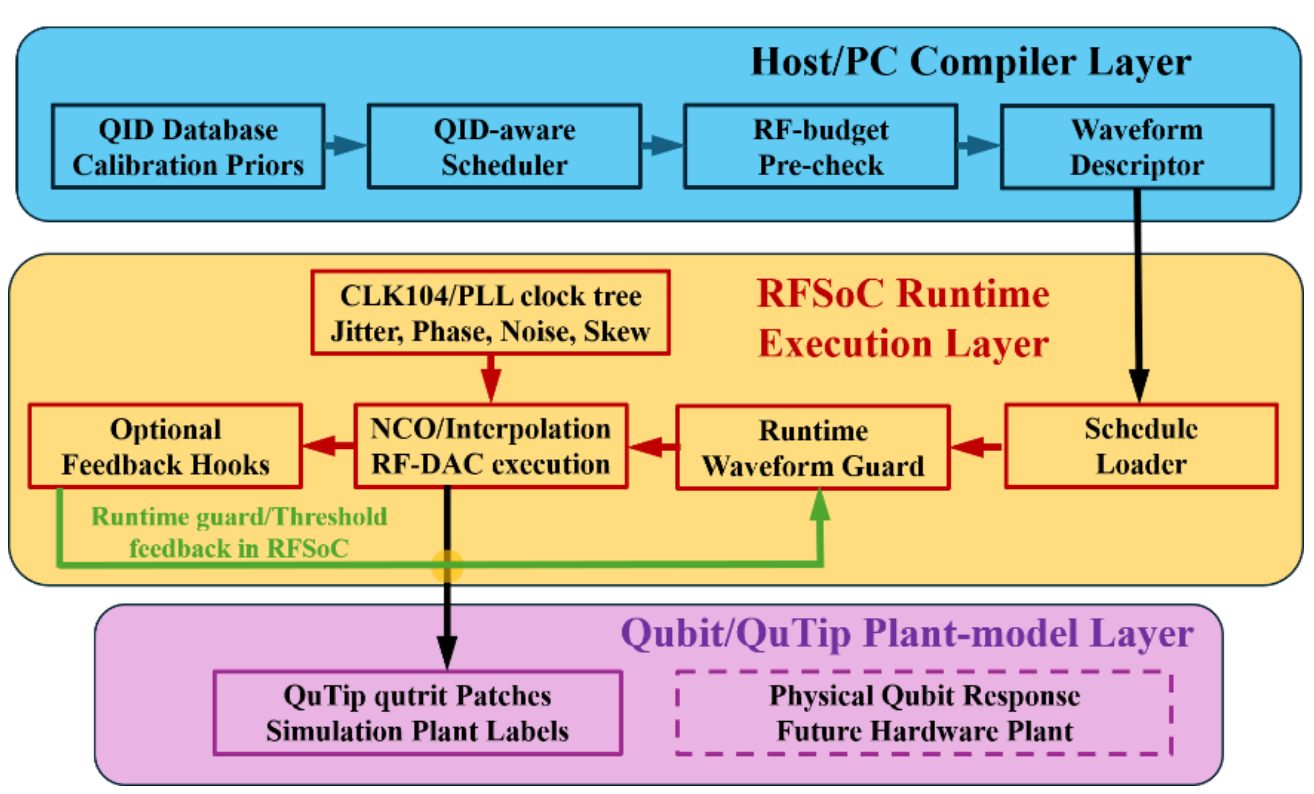

**FIGURE 1. Hardware-software boundary of the QID-aware RF-budgeted workflow. Host-side compilation constructs QID records, schedules RF frames, and emits waveform descriptors. The RFSoC-inspired runtime layer executes the source-chain waveform model, and the plant layer is represented here by local QuTiP qutrit-patch validation.**

### A. Gate Layer, Active Set, and RF-Frame Partition

Let the processor contain the physical-qubit set

$$\mathcal{Q} = \{q_1, q_2, \dots, q_N\}. \quad (1)$$

A requested single-qubit gate layer acts on an active index set $A \subseteq 1, \dots, N$ and is written as

$$\mathcal{G} = \{gate_i = (q_i, \theta_i, \phi_i)\}_{i \in \mathcal{A}}, \quad (2)$$

where $\theta_i$ is the target rotation angle and $\phi_i$ specifies the equatorial rotation axis. For example, X90 corresponds to $\theta_i = \pi/2$ and $\phi_i = 0$, whereas Y90 corresponds to $\theta_i = \pi/2$ and $\phi_i = \pi/2$. The compiler partitions the active set into $K$ RF frames. Every index in the same frame is assigned a tone in one aggregate waveform and is driven during the same frame interval. The primary scheduling objective is to minimize $K$, while validated layer time and crest factor may be used as secondary criteria when several partitions use the same number of frames.

### B. QID Database and QID Record

For shared-line RF-frame compilation, a physical qubit cannot be represented only by a logical index or by its nominal $|0\rangle \leftrightarrow |1\rangle$ carrier frequency. The present formulation requires distinct, resolvable $f_{01}$ frequencies for qubits sharing the same control path and is therefore not applicable when two or more such qubits have identical $f_{01}$ values. The compiler must query qubit-specific transition frequencies, pulse constraints, and drive-scale calibration information when constructing candidate RF frames and evaluating their admissibility. These quantities are organized as a compiler-side QID database, $\mathcal{D}_{\text{QID}} = \{\text{QID}_i\}_{i=1}^{N}$. Each entry is a QID record associated with physical qubit $q_i$,

$$\text{QID}_i = \{f_{01,i}, f_{12,i}, \mathcal{P}_i, \kappa_i\}. \quad (3)$$

here, $f_{01,i}$ is the computational transition frequency, $f_{12,i}$ is the first leakage-transition frequency, $\mathcal{P}_i$ contains the pulse parameters available for qubit $i$, and $\kappa_i$ maps normalized command amplitude to calibrated angular Rabi rate.

The pulse field contains only pulse-shape information. In this work, the pulse family is fixed, and we write

$$\mathcal{P}_i = \{\mathcal{T}_i, \mathbf{e}_i, \beta_i\}. \quad (4)$$

Here, $\mathcal{T}_i$ is the allowed pulse-duration set, $\mathbf{e}_i$ contains the envelope-shape parameters, and $\beta_i$ is the DRAG-like quadrature coefficient. The target rotation angle, rotation-axis phase, commanded carrier frequency, and commanded amplitude are gate-layer or RF-frame quantities and are not stored in $\mathcal{P}_i$.

The stored transition frequencies define the angular-frequency quantities used by the scheduler and decoder, the angular-frequency anharmonicity is a derived compiler quantity,

$$\alpha_i = \omega_{12,i} - \omega_{01,i} = 2\pi\left(f_{12,i} - f_{01,i}\right). \quad (5)$$

Frequency guards are compiler-derived quantities rather than fixed QID fields. For each candidate pair, the compiler constructs a two-tone descriptor and evaluates it using the same RF-chain and transition-frame projection models used for complete frames. The pairwise screen passes only if RF validation passes and the projection diagnostics satisfy the Table II thresholds: intended-drive mismatch $d_{\text{mis}}^{01} \le 5 \times 10^{-3}$, false addressing $d_{\text{FA}}^{01} \le 10^{-3}$, and leakage drive $d_{\text{LD}}^{12} \le 10^{-3}$. The tighter $2 \times 10^{-3}$ mismatch value is retained as a calibration target rather than a hard rejection threshold. Numerically ill-conditioned tone pairs, defined here by a projection Gram-matrix condition number above $10^6$, are also separated.

A separate directed leakage guard protects spectator $|1\rangle \leftrightarrow |2\rangle$ transitions. Its duration-dependent width is calibrated in Section VI-B as the smallest two-sided detuning for which the exact two-qutrit simulation satisfies $P_{2,\max} \le 10^{-3}$. The resulting guard widths are 150, 60, 45, and 30 MHz for pulse durations of 80, 120, 160, and 240 ns, respectively. Intermediate durations use linear interpolation. A pair is separated when either tone falls within the calibrated leakage guard of the other qubit's $f_{12}$ transition.

These pairwise rules are scheduling screens, not final frame-validity claims. Every selected multitone frame is subsequently evaluated using the aggregate RF model, transition-frame projections, and local QuTiP qutrit dynamics so that multitone leakage stacking and other frame-level effects remain visible.

### C. RF-Chain Profile

The RF-chain profile is a separate compiler input and contains RF hardware, DSP, source-chain, and path-response quantities that are not stored in the QID record. We write

$$\mathcal{A}_{\text{RF}} = \{f_s, A_{\text{FS}}, N_{\text{DAC}}, \text{ENOB}, b_{\text{BO}}, \delta u_{\text{eff}}, \mathcal{G}_{\text{DSP}}, B_{\text{DAC}}, \sigma_t, \mathcal{S}_{\text{spur}}, \mathcal{M}_{\text{comp}}, A_{clip}, \mathcal{T}_{src}, \mathcal{H}_{path}\}. \quad (6)$$

Here, $f_s$ is the RF-DAC sampling rate in the configured RFSoC profile, and $A_{\text{FS}}$ is the DAC full-scale amplitude reference used to express the aggregate command waveform. $N_{\text{DAC}}$ and ENOB describe finite converter resolution and ENOB-equivalent noise. The factor $b_{\text{BO}}$ defines the frame-level digital headroom condition for the aggregate multi-tone waveform. The quantity $\delta u_{\text{eff}}$ is an effective normalized command-amplitude resolution floor used for small-angle feasibility checks. It is not an independent chip datasheet parameter. It summarizes DAC quantization, ENOB-equivalent noise, digital gain resolution, and calibration/noise-floor assumptions in the configured RF profile. The set $\mathcal{G}_{\text{DSP}}$ contains the digital timing, frequency, phase, and gain grids used by the waveform sequencer and RFDC/NCO model. These include the sample grid, NCO frequency grid, phase-offset grid, and command-gain grid. $B_{\text{DAC}}$ is the effective DAC/output bandwidth prior, and $\sigma_t$ is the sample-clock jitter prior. The collection $\mathcal{S}_{\text{spur}}$ specifies NCO spurs, DAC spurs, clock feedthrough, and direct-RF artifact terms. The model $\mathcal{M}_{\text{comp}}$ specifies AM-AM and AM-PM compression, and $A_{\text{clip}}$ is the clipping threshold at the node where clipping is evaluated. The operator $\mathcal{T}_{\text{src}}$ maps the descriptor-defined aggregate command waveform to the modeled RF-DAC/source-output waveform. It includes RFDC/NCO quantization, interpolation, finite DAC precision, zero-order-hold response, source bandwidth, clock terms, source-level spurs, compression, and clipping. Because this transformation may be nonlinear, it is applied to the full aggregate frame waveform rather than independently to each tone. The path-response collection $\mathcal{H}_{\text{path}}$ describes the configured transfer from the source output to the qubit-control reference plane. It includes frequency-dependent gain, phase, group delay, filter response, and board/package or control-line $S_{21}$ priors. When path-specific calibration data are unavailable, a shared configured path prior is used.

### D. Compiler Inputs for RF-Frame Construction

The RF-frame construction is specified by the QID database, the requested gate layer, the RF-chain profile, and the effective crosstalk-coupling matrix:

$$\mathcal{I} = \left(\mathcal{D}_{\text{QID}}, \mathcal{G}, \mathcal{A}_{\text{RF}}, \boldsymbol{C}\right), \quad (7)$$

here, $\mathcal{D}_{\text{QID}}$ provides qubit-specific transition, pulse, and drive-calibration information; $\mathcal{G}$ specifies the requested active single-qubit rotations; $\mathcal{A}_{\text{RF}}$ defines the configured RF-chain and path-response model; and $\mathbf{C}$ is the effective crosstalk-coupling matrix. Its element $\mathrm{C}_{\mathrm{ji}}$ denotes the effective coupling of tone $\mathrm{i}$ into qubit $\mathrm{j}$, with the normalization $\mathrm{C}_{\mathrm{ii}} = 1$. In the shared-line limit, $\mathrm{C}_{\mathrm{ji}} = 1$ for all relevant $\mathrm{i}, \mathrm{j}$; partial isolation is represented by $0 < \mathrm{C}_{\mathrm{ji}} < 1$.

These inputs are used in Section III to synthesize candidate multitone RF frames. Frame acceptance, repartitioning, and local validation are then defined by the admission procedures in Sections IV and V.

## III. RF-FRAME ENCODING AND SOURCE-CHAIN RECONSTRUCTION

### A. Calibrated Single-Tone Envelope

For a target rotation angle $\theta_i$, the pulse-area condition is

$$\theta_i = \int_0^{T_i} \Omega_i(t)\,dt = \kappa_i u_i \int_0^{T_i} g_i(t)\,dt = \kappa_i u_i I_i, \quad (8)$$

where $\Omega_i(t)$ is the calibrated angular Rabi rate applied to qubit $q_i$. The selected in-phase pulse envelope is denoted by $g_i(t)$ and is peak-normalized unless otherwise stated. $I_i$ is the pulse-area factor, $\kappa_i$ maps normalized command amplitude to angular Rabi rate, and $u_i$ is the calibrated normalized command amplitude assigned to tone $i$. Thus,

$$u_i = \frac{\theta_i}{\kappa_i I_i}. \quad (9)$$

The complex baseband envelope assigned to tone $i$ is

$$d_i(t) = u_i e^{j\phi_i}\left[g_i(t) - j\beta_i \frac{\dot{g}_i(t)}{\alpha_i}\right], \quad (10)$$

where $\phi_i$ is the target rotation-axis phase, $\beta_i$ is the configured quadrature weight, and $\alpha_i$ is the anharmonicity convention used by the pulse calibration. If the stored anharmonicity is expressed in cyclic-frequency units, the corresponding $2\pi$ convention is absorbed into the configured value of $\beta_i$.

### B. Aggregate RF-Frame and Source-Chain Transformation

For candidate frame $F$, each calibrated complex envelope $d_i(t)$ is modulated onto its assigned carrier frequency $f_i$, and the active tones are summed to form the physical real RF command waveform:

$$s_F(t) = \mathrm{Re}\{\textstyle\sum_{i\in F} d_i(t) e^{j2\pi f_i t}\}. \quad (11)$$

The complete aggregate waveform is then propagated through the configured source-chain model,

$$s_{F,\mathrm{src}}(t) = \mathcal{H}_{\mathrm{src}}[s_F(t)]. \quad (12)$$

The operator $\mathcal{H}_{\mathrm{src}}$ represents the modeled RFDC interpolation, NCO and DAC quantization, finite DAC precision, zero-order-hold response, source bandwidth, clock terms, source-level spurs, compression, clipping, and RF-path response. Because this transformation may be nonlinear, it is applied to the complete aggregate frame rather than independently to each tone. In the implemented workflow, the encoded and propagated RF command carries the in-phase envelope $g_i(t)$; the DRAG-like quadrature term of $d_i(t)$ is applied at the decoded qutrit-drive stage, so the headroom and source-chain checks act on the in-phase aggregate.

## IV. SCHEDULING AND RF VALIDATION

QID-aware RF control cannot assume that all logically simultaneous gates can be executed in a single RF frame. Gates may need to be separated when the combined QID, RF-profile, and effective-coupling inputs indicate carrier crowding, leakage-transition proximity, strong crosstalk, or RF-chain incompatibility. This section describes the compiler-side workflow used to construct the conflict graph, partition an active gate layer into candidate RF frames, form the corresponding descriptors, and evaluate them against the configured RF-validation conditions. A failed candidate either adds a no-good constraint and triggers recoloring or is rejected when no RF-valid partition can be found under the configured RF profile.

Figure 2 summarizes the implemented scheduling and RF-validation workflow. The active gate layer is converted into a QID-level conflict graph and partitioned using the exact coloring procedure of Ref. [35]. For each candidate coloring, the compiler forms RF-frame descriptors and evaluates them against the Table-I RF-validation conditions. A re-partitionable failure adds a no-good constraint and triggers recoloring. If no coloring produces a complete RF-valid partition, the requested layer is classified as hardware-limited under the configured RF profile. Only RF-valid descriptors are emitted to the RFSoC runtime.

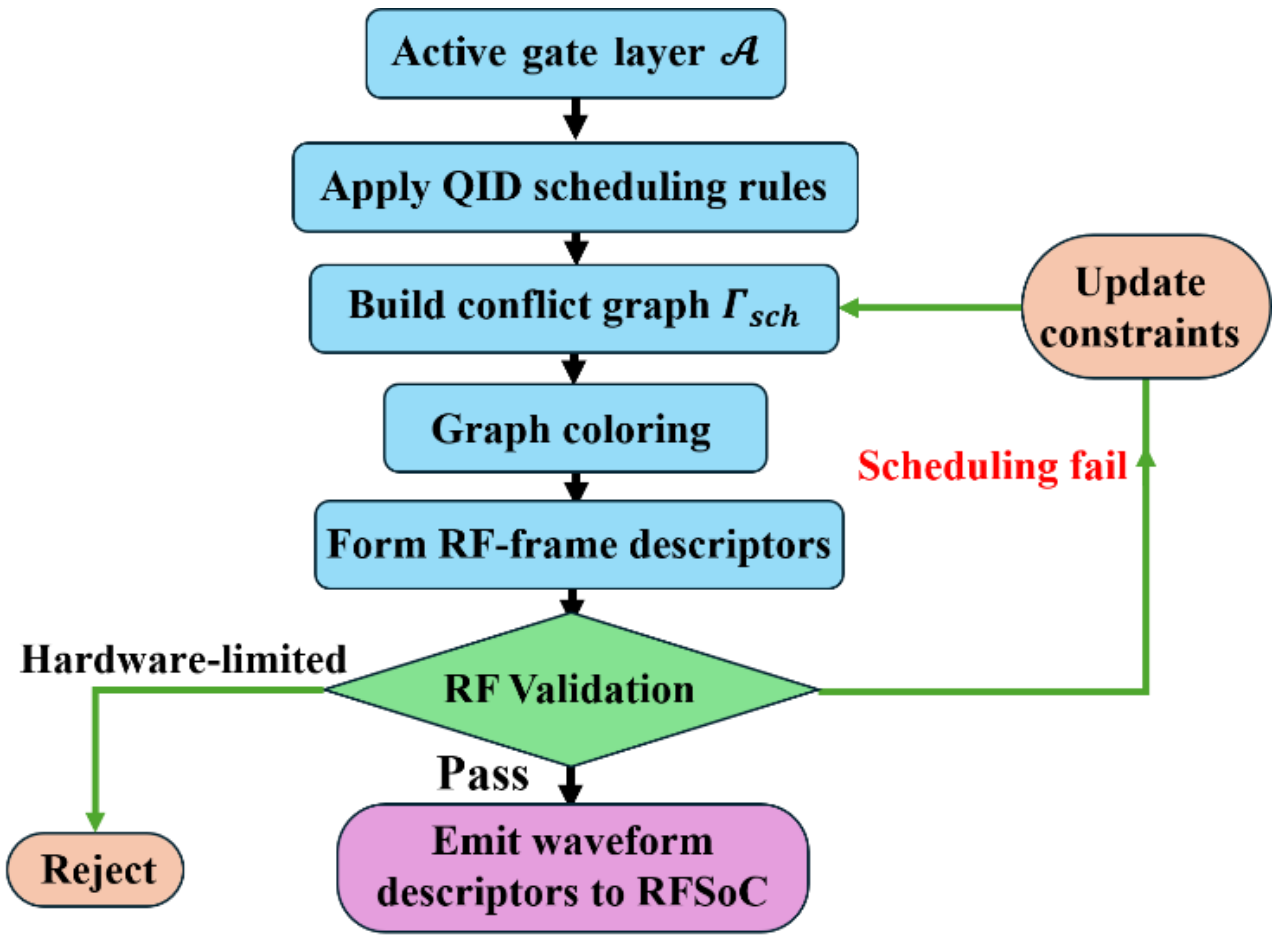

FIGURE 2. **RF-frame scheduling and RF-validation workflow. Exact graph coloring forms candidate RF-frame descriptors, while RF-validation failures add no-good constraints and trigger recoloring. Only complete RF-valid partitions are emitted to the RFSoC runtime; otherwise, the layer is classified as hardware-limited.**

### A. QID Scheduling Rules

The scheduler evaluates every candidate pair of logically simultaneous gates before graph coloring. Rather than applying a single carrier-spacing cutoff, it forms the corresponding two-tone descriptor, propagates it through the configured RF model, and applies three explicit coexistence rules.

First, the pair must pass the RF-admission checks for aggregate headroom, DAC-input clipping, and the command-amplitude floor. Second, the two $f_{01}$ tones must remain distinguishable and must not produce excessive off-target $|0\rangle \leftrightarrow |1\rangle$ drive. This condition is evaluated from the post-RF tone projections: intended-drive mismatch, false addressing, and leakage-drive projection must satisfy the Table II limits of $5\times10^{-3}$, $1\times10^{-3}$, and $1\times10^{-3}$, respectively. The effective coupling coefficients $C_{ji}$ are included in these projections. The joint tone-extraction condition number must also remain below $10^{6}$.

Third, both directed leakage conditions are checked explicitly: tone $i$, located at $f_{01,i}$, must remain outside the guarded neighborhood of $f_{12,j}$, and tone $j$ must remain outside that of $f_{12,i}$. The calibrated guard half-widths used by the scheduler are 150, 60, 45, and 30 MHz for pulse durations of 80, 120, 160, and 240 ns, respectively. Thus, closely spaced $f_{01}$ carriers create an edge when their spectral overlap causes excessive false addressing or poor tone separation, whereas an $f_{01}$ carrier near another qubit's $f_{12}$ transition creates an edge through the directed leakage guard. An edge is inserted if any one of these tests fails. No separate heuristic collision-score threshold is used.

### *B. Conflict Graph Construction and Graph Coloring*

The pairwise incompatibility relation from Sec. IV-A is converted into a scheduling conflict graph. Each active gate is represented as one vertex, and an edge is inserted when two gates cannot share the same RF frame:

$$\Gamma_{\mathrm{sch}} = (\mathcal{A}, E_{\mathrm{sch}}), \quad (13)$$

$$(i,j) \in E_{\mathrm{sch}} \Leftrightarrow \mathrm{Conflict}(i,j) = 1. \quad (14)$$

A valid RF-frame schedule is a vertex coloring of this graph. Each gate $i$ is assigned a frame index $c_i$, and conflicting gates must receive different indices:

$$c_i \neq c_j, \ \forall (i,j) \in E_{\mathrm{sch}}. \quad (15)$$

Each color class defines one candidate RF frame:

$$F_k = \{i \in \mathcal{A} : c_i = k\} \quad (16)$$

The candidate schedule is therefore

$$\Pi_{\mathrm{sch}} = \{F_1, \dots, F_K\}. \quad (17)$$

In the reported implementation, the coloring step uses an exact DSATUR-style branch-and-bound search. DSATUR selects the next vertex using saturation degree, i.e., the number of distinct colors already present among its colored neighbors, with graph degree used as a tie-breaker. This ordering was introduced by Brélaz for graph coloring and is used here inside an exact $K$-colorability search. [35]

The solver tests increasing frame counts until it finds the smallest feasible coloring of the constructed conflict graph:

$$K_{\mathrm{sch}}^{\star} = \min K \text{ s.t. } \exists \{c_i\}_{i\in\mathcal{A}} \text{ with } c_i \neq c_j \ \ \forall (i,j) \in E_{\mathrm{sch}}. \quad (18)$$

Thus, the scheduling stage returns a minimum-frame partition with respect to the QID-level pairwise conflict graph. This minimum is not yet a final hardware-admissibility claim: model-based validation may later reject a candidate frame, add a no-good constraint, and trigger recoloring.

### *C. RF-Frame Descriptor Formation*

After graph coloring, each color class defines one candidate RF frame. The compiler then converts each frame into an RF-frame descriptor. This descriptor is a parameter object for downstream validation and RFSoC execution; it is not a sampled physical waveform.

For a candidate frame $F_k$, the descriptor can be written compactly as

$$\mathbf{d}_{F_k} = \{q_i, f_i, u_i, \phi_i, p_i, T_i, t_i\}_{i\in F_k}. \quad (19)$$

here $q_i$ is the addressed qubit, $f_i$ is the assigned carrier frequency, $u_i$ is the calibrated amplitude or gain setting, $\phi_i$ is the programmed phase, $p_i$ specifies the pulse family or envelope parameters, $T_i$ is the pulse duration, and $t_i$ is the timing offset within the frame. Frame-level metadata may also include the RF output channel, reference frequency, RF-profile identifier, and operating constraints such as backoff or tone-count limits.

This separation is important. The scheduler determines which gates share a frame and forms the corresponding descriptors. The RFSoC runtime later uses these descriptors to synthesize the actual waveform. Likewise, model-based validation may instantiate a predicted waveform internally from the descriptor, but the compiler output at this stage remains a descriptor-level representation.

### *D. RF Frame Validation*

For each candidate RF frame, the waveform descriptor is instantiated by the RFSoC-informed behavioral runtime. The descriptor configures the Zynq PS control registers and frame sequencer, while the PL tone-generator bank synthesizes each active tone from its assigned carrier frequency, phase, calibrated gain, and envelope. The active tones are summed into one aggregate digital stream. Descriptor-level headroom is checked before AXI4-Stream transfer. The RFDC then applies the configured 2× interpolation, after which clipping is evaluated at the DAC input and the 14-bit RF-DAC/source model generates the output waveform. The CLK104/PLL block supplies the 6.144-GS/s sampling clock and clock-jitter prior. The PS, AXI-Lite, and AXI4-Stream blocks identify control and data-transfer boundaries; bus timing is not modeled. This descriptor-to-RF-DAC runtime path is summarized in Figure 3(a).

After RF-DAC conversion, the aggregate waveform propagates through the configured source-to-qubit path. The modeled shared-path response $\mathcal{H}_{\text{path}}(f)$ represents the combined transfer through the RF output path, the SMA/coax interconnect, the SMA-to-coplanar-waveguide (CPW) launch, and the shared CPW control line to the qubit-control reference plane. The resulting waveform is delivered to multiple frequency-distinct transmons connected to the same control line, as illustrated in Figure 3(b).

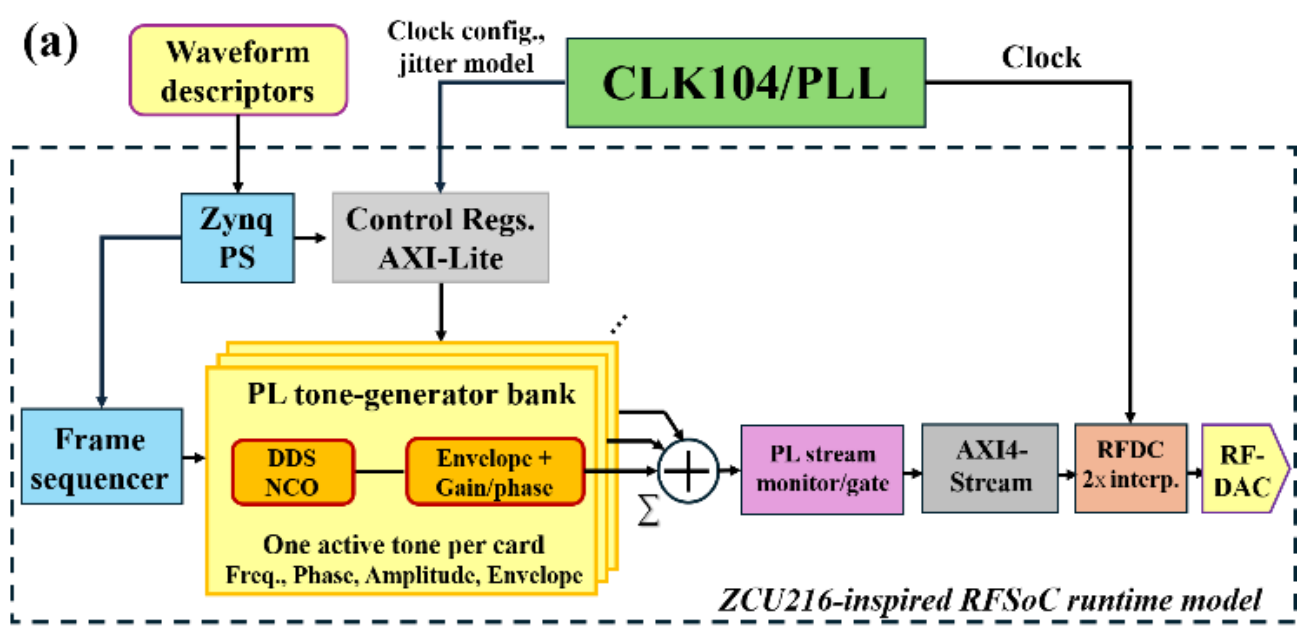

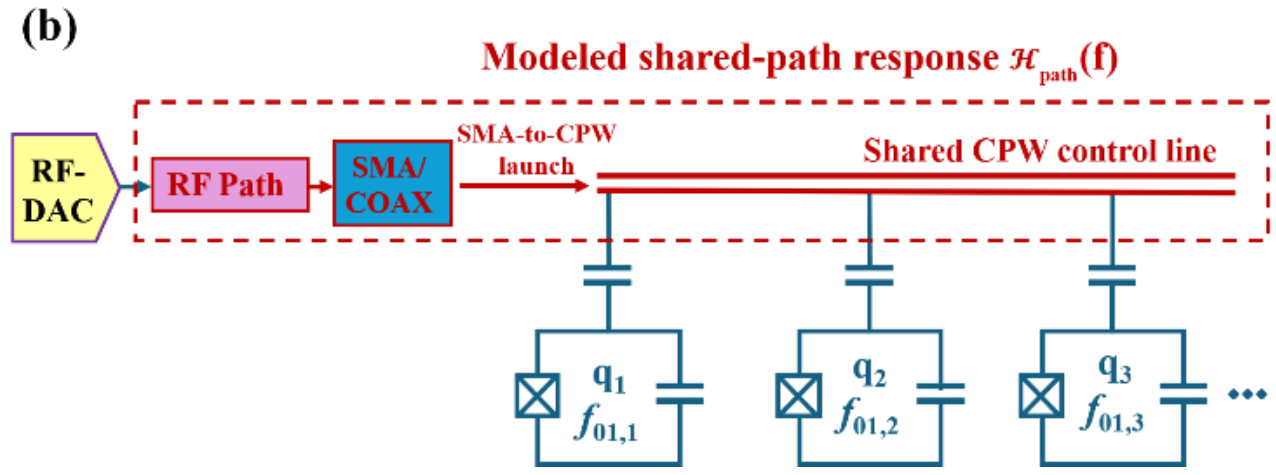

**FIGURE 3. RFSoC-informed runtime and shared-line interface represented by the behavioral model. (a) Descriptor-driven multitone synthesis, stream monitoring, RFDC interpolation, and RF-DAC conversion. (b) Modeled shared-path response $\mathcal{H}_{\text{path}}(f)$ from the RF-DAC output through the RF path and shared CPW control line to frequency-distinct transmons.**

Table I lists the configured RF-budget channels used in the RFSoC-informed behavioral validation model. Device-level RFSoC capabilities, including the ZCU216/ZU49DR platform, converter resolution, RFDC interpolation/NCO functionality, and CLK104 clocking architecture, are based on AMD documentation [36]-[40]. RFSoC Gen3 characterization literature is used only as external context for plausible source-chain impairments [41]. The remaining numerical values marked as priors or stress brackets are configured simulation settings and should not be interpreted as manufacturer-guaranteed ZCU216 specifications.

**TABLE I. Configured RF-budget channels used in the RFSoC-informed behavioral validation model.**

| Runtime stage | Budget channel | Model value used in this study | Budget / threshold |
|---|---|---|---|
| Frame timing | Sample rate | $f_s = 6.144$ GS/s*; $f_{\text{desc}} = 3.072$ GS/s* | Descriptor timing must lie on the $f_{\text{desc}}$ grid; RF-DAC-rate samples lie on the $f_s$ grid. |
| RFDC interpolation | Interpolation model | $L = 2$*; 30th-order Blackman-windowed sinc FIR* | Included in $\mathcal{T}_{\text{src}}$; no independent pass/fail threshold. |
| PL tone generator | Per-tone DDS control grid | $N_{f,PL} = 48$ bits*; $N_{\Phi,PL} = 48$ bits* | Per-tone frequency offsets and phases are quantized to the configured PL grids; no independent pass/fail threshold. |
| RFDC fine mixer | Common carrier placement | study-dependent $f_{\text{ref}}^*$; $N_{f,\text{RFDC}} = 48$ bits; $N_{\Phi,\text{RFDC}} = 18$ bits; equivalent-complex representation* | One common translation is applied to the aggregate frame, with $f_{01,k} = f_{\text{ref}} + \Delta f_k$ and $f_{01,k} <$ 6.0 GHz*. No second per-tone quantization is applied. |
| PL tone generator | Minimum useful amplitude | $\delta u_{\text{eff}} = 5 \times 10^{-4}$ FS†;$2 \times 10^{-3}$ FS‡ | Pass if every requested tone amplitude is at least $\delta u_{\text{eff}}$ |
| PL multitone sum | peak headroom | $A_{\text{FS}} = 1$ FS, $b_{\text{BO}} = 10^{-1/20} = 0.891$ | pass if $max_t \mid s_F(t) \mid \leq b_{BO} A_{FS}$ |
| PL multitone sum | PAPR / crest factor | computed from each frame waveform | reported diagnostic only; headroom/clipping provide pass/fail checks |
| Stream monitor / DAC input | clipping level | $A_{\text{clip}} = 1$ FS*; 0.60/0.45 FS‡ stress bracket: 0.60 FS / 0.45 FS | pass if $\max_t \mid s_{\text{DAC,in}}(t) \mid < A_{\text{clip}}$ |
| CLK104 / PLL | clock jitter | $\sigma_t = 250$ fs rms†; 5/20 ps‡ | included as jitter-induced phase/amplitude error |
| RF-DAC model | DAC resolution | $N_{\text{DAC}} = 14$ bits | Finite-resolution I/Q quantization |
| RF-DAC model | ENOB-equivalent noise | ENOB = 11.5 bits† | $\sigma_q = 2^{-\text{ENOB}}/\sqrt{12} \approx 1.0 \times 10^{-4}$ $\text{FS}_{\text{rms}}$ |
| RFDC / DAC response | ZOH and bandwidth | First-order ZOH*; $B_{\text{DAC}} = 2.0$ GHz†; 1.8/ 1.2 GHz‡ | $H_B(f) = \exp\left[-\frac{12}{(f/B_{\text{DAC}})}^4\right]$; pass/fail is workload dependent and determined by Table II closure tests. |
| RF-DAC artifacts | DAC spur | −65 dBc† at +90 MHz* | Full-scale-referenced additive artifact |

| Runtime stage | Budget channel | Model value used in this study | Budget / threshold |
|---|---|---|---|
| DDS/NCO artifacts | NCO spur | $-80$ dBc† at +35 MHz* | Full-scale-referenced additive |
| Output RF network | AM-AM / AM-PM compression | $\alpha_{\text{comp}} = 0^{\dagger}$, $\beta_{\text{AMPM}} = 0^{\dagger}$; $\alpha_{\text{comp}} = 0.3/0.6^{\ddagger}$ | Applied to the complete aggregate waveform; closure is workload dependent and evaluated using Table II. |
| Output RF network | group-delay ripple | 0 ps†;50 ps‡; modeled period 2.5 GHz* | Included in $\mathcal{H}_{\text{path}}$; no independent threshold. |
| Output RF network | path transfer response | $S_{21} = (-1.5, 0, -1.5)$ dB† at offsets $(-2.5, 0, +2.5)$ GHz* | Included in static per-tone calibration $\gamma_k$ and in the post-RF response; residual errors are evaluated using Table II. |

∗ Study-specific configuration or behavioral-model choice. † Circuit-informed prior not measured on the referenced ZCU216/CLK104 hardware. ‡ Configured sensitivity or stress value. Only the $\delta u_{\text{eff}}$ sweep is exercised in the reported results; the other stress brackets are configured but not exercised. Unmarked values are public device characteristics or quantities computed directly from each simulated frame.

The common carrier reference is $f_{\text{ref}} = 5.0$ GHz for the single-qutrit study, the midpoint of the two $f_{01}$ values for the pairwise study, and 5.25 GHz for the multiqubit-capacity and algorithm-derived-workload studies.

The 4.75–5.75-GHz carrier band is represented as an equivalent complex envelope about the study-specific $f_{\text{ref}}$ listed in Table I and is interpreted within a sub-6-GHz direct-RF/mix-mode operating region.

For each candidate RF frame, the descriptor-defined aggregate waveform is checked against the configured RF-budget channels in Table I without post-hoc renormalization of the calibrated tone amplitudes. A frame is RF-admitted only if all hard admission checks pass. PAPR is reported only as a crest-factor diagnostic unless an explicit PAPR limit is configured. Frames that fail RF validation are rejected before transition-frame decoding and local qutrit-patch simulation.

## V. DECODING AND FIDELITY

After a candidate RF-frame descriptor passes RF-budget validation, it is propagated through the RFSoC-informed source-chain model and the configured source-to-qubit path model. The resulting effective drives are then decoded into transition-frame quantities and evaluated by local qutrit-patch simulation. The quantities in this section are simulation diagnostics.

### A. Transition-Frame Projection and Screening Coefficients

For each RF-admitted frame, the implementation first recovers the complex post-RF gain of every tone by a joint least-squares projection of the aggregate waveform onto the interpolated single-tone bases. For each tone-target pair, it then evaluates the envelope-area-normalized complex spectral overlap at the target qubit's $f_{01}$ and $f_{12}$ transitions. Each overlap is multiplied by the actual-to-requested command-amplitude ratio, the recovered complex tone gain, and the effective coupling $C_{ji}$; the leakage term additionally includes the transmon $\sqrt{2}$ matrix-element factor. The intended-drive mismatch is the deviation of the addressed $f_{01}$ coefficient from unity. False addressing is obtained by coherently summing all off-target $f_{01}$ contributions, whereas leakage drive is obtained by coherently summing all $f_{12}$ contributions, including the addressed tone. The frame-level quantities $d_{\text{mis}}^{01}$, $d_{\text{FA}}^{01}$, and $d_{\text{LD}}^{12}$ are the maximum magnitudes over addressed qubits and are compared with the Table II thresholds. These quantities are screening diagnostics rather than gate fidelities; admitted frames are subsequently evaluated by local QuTiP qutrit-patch simulation.

### B. Rotation Reconstruction and Angle Accuracy

For each RF-admitted frame, the local qutrit-patch QuTiP simulation returns the computational-subspace map $U_{j,comp}(F)$ for each addressed qubit $j$. The effective rotation angle $\hat{\theta}_j(F)$ and equatorial-axis phase $\hat{\Phi}_j(F)$ are obtained from the Bloch-vector expectation values of the survival-projected final state after evolving qubit j from $|0\rangle$ under all frame tones. The absolute rotation-angle error and the principal-value axis-phase error, wrapped modulo $2\pi$, are evaluated relative to their target values. The frame-level metrics are the maximum values over all addressed qubits, as summarized in Table II. The axis-phase error is evaluated only when the final-state azimuth is well defined; it is masked for target rotations that place the Bloch vector near a pole.

### C. Fidelity and Leakage Diagnostics

For each RF-admitted frame $F$, the decoded post-RF tones are applied to local qutrit patches. In the rotating frame of the computational transition of qutrit $j$, with $\hbar = 1$, the driven three-level Hamiltonian is

$$H_{j,F}(t) = 2\pi\alpha_j |2_j\rangle\langle 2_j| + \tfrac{1}{2}\sum_{i\in F}\left[C_{ji}\tilde{d}_i(t)e^{-i2\pi(f_i - f_{01,j})t}a_j^{\dagger} + \text{H.c.}\right], \quad (20)$$

where $\alpha_j = f_{12,j} - f_{01,j}$, $C_{ji}$ is the effective coupling from tone $i$ to qutrit $j$, and

$$a_j = |0_j\rangle\langle 1_j| + \sqrt{2}\,|1_j\rangle\langle 2_j|. \quad (21)$$

The decoded envelope $\tilde{d}_i(t)$ includes the post-RF complex tone gain and the DRAG quadrature defined in Section III. The RF command propagated through the source chain contains the in-phase envelope; the DRAG quadrature is introduced at the decoded qutrit-drive stage. The Hamiltonian of a selected patch is the sum of these local driven-qutrit Hamiltonians. Direct inter-qutrit coupling and $T_1/T_2$ dissipation are not included in the present simulations.

For each addressed qutrit, propagation from the computational basis states $|0\rangle$ and $|1\rangle$ constructs the projected computational-subspace map $K_j = P_{01}U_{j,F}P_{01}$. Its average computational-subspace survival is $S_{\mathrm{comp},j} = \mathrm{Tr}(K_j^{\dagger}K_j)/2$. The leakage-aware average gate fidelity relative to the target gate $U_j^{\mathrm{tar}}$ is

$$F_{\mathrm{avg},j}^{\mathrm{leak}} = \frac{\mathrm{Tr}\left(K_j^{\dagger}K_j\right)+\left|\mathrm{Tr}\left[\left(U_j^{\mathrm{tar}}\right)^{\dagger}K_j\right]\right|^2}{6}. \quad (22)$$

This quantity penalizes both loss from the computational subspace and coherent mismatch with the target operation.

A separate polar-unitary coherent-fidelity diagnostic is obtained from the unitary factor $V_j$ in the polar decomposition $K_j = V_jP_j$. It is evaluated as

$$F_{\mathrm{PU},j} = \left(\left|\mathrm{Tr}\left[\left(U_j^{\mathrm{tar}}\right)^{\dagger}V_j\right]\right|^2 + 2\right)/6. \quad (23)$$

Because this construction removes the nonunitary part of the projected map, $F_{\mathrm{PU}}$ is reported only as a coherent-error diagnostic and does not enter the Table II closure decision.

Maximum transient leakage is the largest $|2\rangle$-state population reached during propagation. For pulse-shape diagnostics, the terminal leakage population at the end of the frame is also reported; it is diagnostic only and does not enter the Table II closure decision. The time-resolved trajectory additionally provides the minimum instantaneous computational-subspace population, which is kept distinct from the map-based $S_{\mathrm{comp}}$.

At frame level, $F_{\mathrm{avg}}^{\mathrm{leak}}(F)$, $F_{\mathrm{PU}}(F)$, and $S_{\mathrm{comp}}(F)$ are the minimum values over the simulated addressed qutrits, whereas $P_2^{\max}(F)$ is the maximum over qutrits and propagation time. Terminal leakage, when shown, is likewise reported as a worst-case diagnostic. These quantities characterize closure under the modeled Hamiltonian and RF chain; they are not experimentally measured process or randomized-benchmarking fidelities.

**TABLE II. Configured frame-closure thresholds used in this study.**

| Closure channel | Diagnostic quantity | Threshold |
|---|---|---|
| Intended-drive mismatch | $d_{\mathrm{mis}}^{01}(F)$ | $\tau_{\mathrm{mis}} = 5\times10^{-3}$ |
| Computational false addressing | $d_{\mathrm{FA}}^{01}(F)$ | $\tau_{\mathrm{FA}} = 1\times10^{-3}$ |
| Leakage-drive projection | $d_{\mathrm{LD}}^{12}(F)$ | $\tau_{\mathrm{LD}} = 1\times10^{-3}$ |
| Rotation-angle error | $\Delta\theta(F)$ | $\tau_{\theta} = 1^{\circ}$ |
| Rotation-axis phase error | $\Delta\phi(F)$ | $\tau_{\phi} = 1^{\circ}$ |
| Leakage-aware fidelity loss | $1 - F_{\mathrm{avg}}^{\mathrm{leak}}(F)$ | $\tau_{F,\mathrm{leak}}=1\times10^{-3}$ |
| Polar-unitary coherent-fidelity loss | $1 - F_{\mathrm{PU}}(F)$ | Diagnostic only |
| Computational-survival loss | $1 - S_{\mathrm{comp}}(F)$ | $\tau_S = 1\times10^{-3}$ |
| Maximum transient leakage | $P_2^{\max}(F)$ | $\tau_{P2} = 1\times10^{-3}$ |

**Reproducibility**. The MATLAB workflows were executed in MATLAB R2025a Update 1. The QuTiP bridge used Python 3.12.13 and QuTiP 5.3.0. Local qutrit dynamics were propagated with QuTiP's sesolve using the Adams integrator **atol** = $\mathbf{1\times10^{-8}}$, **rtol** = $\mathbf{1\times10^{-6}}$, and **nsteps** = **2500**. The single-qutrit, pairwise, multiqubit-capacity, and algorithm-derived-workload studies used base pseudorandom seeds 2100, 2202, 3303, and 4404, respectively, with deterministic setting- and case-dependent offsets specified in the accompanying scripts.

## VI. EXPERIMENTS AND RESULTS

All results in this section use the configured RF/QID/QuTiP simulation workflow. We report both frame-normalized and time-normalized layer aggregation. The frame-normalized aggregation is

$$\rho_{\mathrm{layer}} = \frac{N_{\mathrm{active}}}{KB_{\mathrm{RF}}} \quad [\mathrm{qubits/GHz/frame}], \quad (24)$$

where $N_{\mathrm{active}}$ is the number of active microwave rotations, $K$ is the number of validated RF frames required to cover the layer, and $B_{\mathrm{RF}}$ is the RF bandwidth occupied by the QID map.

The validated layer time is

$$T_{\mathrm{layer}} = \sum_{f=1}^{K} T_f. \quad (25)$$

The corresponding time-normalized aggregation is

$$\rho_{\mathrm{time}} = \frac{N_{\mathrm{active}}}{B_{\mathrm{RF}}T_{\mathrm{layer}}} \quad [\mathrm{qubits/(GHz\cdot\mu s)}]. \quad (26)$$

In the present experiments, all frames in a tested layer use the same pulse duration $T$, so $T_{\mathrm{layer}} = KT$. Thus, $\rho_{\mathrm{layer}}$ measures the average number of admitted tones per unit RF bandwidth and frame, whereas $\rho_{\mathrm{time}}$ additionally accounts for the pulse-duration cost. Both quantities are reported only for complete validated partitions in which every frame passes the configured RF and closure criteria.

### A. Single-Qutrit Pulse-Closure Baseline

Before evaluating pairwise or multitone frames, we establish a single-tone closure baseline for one calibrated transmon-like qutrit. This experiment isolates RF admission, pulse-shape error, and leakage before pairwise interference or scheduling is introduced.

The benchmark qutrit has $f_{01} = 5.00$ GHz, $f_{12} = 4.75$ GHz, and anharmonicity $\alpha = -250$ MHz. A resonant X-axis drive uses a truncated Gaussian envelope with $\sigma/T = 0.18$ and the DRAG-like quadrature defined in Section III-A. Unless $\beta$ is explicitly swept, $\beta = 0.5$ is used. The calibrated command reference is $0.25A_{\mathrm{FS}}$ for an $X90$ rotation at $T = 120$ ns. The requested peak amplitude is therefore

$$u_{\mathrm{pk}} = 0.25A_{\mathrm{FS}}\left(\frac{\theta}{90^\circ}\right)\left(\frac{120\ \mathrm{ns}}{T}\right). \quad (27)$$

The angle-duration scan contains 54 combinations with

$$\theta \in \{1^\circ, 2^\circ, 5^\circ, 10^\circ, 20^\circ, 45^\circ, 90^\circ, 135^\circ, 180^\circ\}$$

and

$$T \in \{20, 40, 80, 120, 160, 240\}\ \mathrm{ns}.$$

Each command is encoded without post-assignment renormalization, propagated through the configured RF chain, and evaluated using the Table II closure criteria and exact single-qutrit QuTiP dynamics.

Of the 54 angle-duration settings in Figure 4, 49 pass RF admission and 41 satisfy all applicable closure criteria. The red-outlined white regions identify the five settings rejected by RF validation before QuTiP evaluation: X90 at 20 ns, X135 at 20 and 40 ns, and X180 at 20 and 40 ns. In each case, the requested waveform exceeds the configured aggregate-headroom limit.

Figure 4(a) reports the leakage-aware fidelity loss. Among the RF-admitted settings, its maximum value is $1.56 \times 10^{-6}$, well below the $10^{-3}$ threshold in Table II. Figure 4(b) shows that pulse duration more strongly affects maximum transient leakage. For X180, $P_2^{\max}$ decreases from $9.74 \times 10^{-4}$ at 80 ns to $4.33 \times 10^{-4}$ at 120 ns and $1.08 \times 10^{-4}$ at 240 ns. X180 therefore first closes at 80 ns, with only a 2.6% margin below the $10^{-3}$ transient-leakage threshold. X90 closes from 40 to 240 ns, with $P_2^{\max}$ decreasing from $3.26 \times 10^{-4}$ to $9.01 \times 10^{-6}$.

Figure 4(c) reports rotation-angle error; the largest value among the RF-admitted settings is $0.126^\circ$, below the $1^\circ$ threshold. Figure 4(d) reports rotation-axis phase error only when the final-state azimuth is well defined. The black-dashed regions identify the near-polar targets for which this diagnostic is not reported; in the present grid these are $1^\circ$, $2^\circ$, $5^\circ$, and $180^\circ$. The largest reported phase error is $0.069^\circ$, also below its $1^\circ$ threshold. The remaining eight RF-admitted but nonclosing settings fail either the intended-drive-mismatch or leakage-drive-projection criterion. In particular, at $\theta = 1^\circ$, the 80, 160, and 240 ns settings close; the 20 ns setting fails the leakage-drive projection, while the 40 and 120 ns settings fail intended-drive mismatch. This nonmonotonic structure arises because full-scale-referenced RF artifacts and noise become comparable to the very small command and should not be interpreted as a monotonic minimum-angle boundary.

Figure 5 separates the pulse-shaping results for the X90 and X180 targets. Panels (a)–(c) show the X90 maximum transient leakage, final leakage, and leakage-aware fidelity loss, respectively; panels (d)–(f) show the corresponding X180 quantities. The minimum-useful-amplitude floor is disabled in these scans so that the colored regions isolate pulse-shaping behavior. The white regions enclosed by red dashed lines are rejected by RF validation before QuTiP evaluation. Under the configured headroom constraint, X90 is admitted for $T \geq 40$ ns, whereas X180 requires $T \geq 80$ ns.

For every RF-admitted duration, $\beta = 0.5$ minimizes the leakage-aware fidelity loss. At $T = 120$ ns, the X90 loss decreases from $1.57 \times 10^{-5}$ at $\beta = 0$ to $9.28 \times 10^{-8}$ at $\beta = 0.5$; the corresponding X180 loss decreases from $1.64 \times 10^{-4}$ to $6.63 \times 10^{-7}$. By contrast, the maximum transient leakage depends primarily on pulse duration and only weakly on $\beta$. At $\beta = 0.5$ and $T = 120$ ns, $P_{2,\max} = 3.60 \times 10^{-5}$ for X90 and $4.33 \times 10^{-4}$ for X180, while the final leakage falls to $3.93 \times 10^{-8}$ and $3.10 \times 10^{-7}$, respectively. This separation shows that most of the level-$|2\rangle$ population is transient and returns by the end of the pulse. The closure test nevertheless retains $P_{2,\max}$ as the conservative leakage diagnostic.

Figure 6 isolates the minimum-useful-amplitude branch of RF validation. A command is rejected by the RF guard when its requested per-tone peak amplitude falls below $\delta u_{\mathrm{eff}}$; no amplitude rounding or renormalization is applied. At the nominal $\delta u_{\mathrm{eff}} = 5 \times 10^{-4}$ FS, the sampled minimum admitted target is $0.25^\circ$ for $T = 120$ ns and $0.5^\circ$ for $T = 240$ ns. Increasing the pulse duration reduces the peak amplitude required for a fixed rotation area and therefore makes small-angle commands more likely to fall below the configured amplitude floor. These regions describe RF-floor admission only; admitted commands must still satisfy the remaining Table II closure criteria.

QuTiP output-grid convergence was checked for $X90$ at 120 ns and $X180$ at 80 and 120 ns using 400, 800, and 1600 output samples. Between the 800- and 1600-sample results, the largest changes were $2.35 \times 10^{-4}$ degrees in rotation-angle error, $1.19 \times 10^{-7}$ in maximum transient leakage, and $2.18 \times 10^{-9}$ in leakage-aware fidelity loss. All three changes are below the prescribed convergence tolerances.

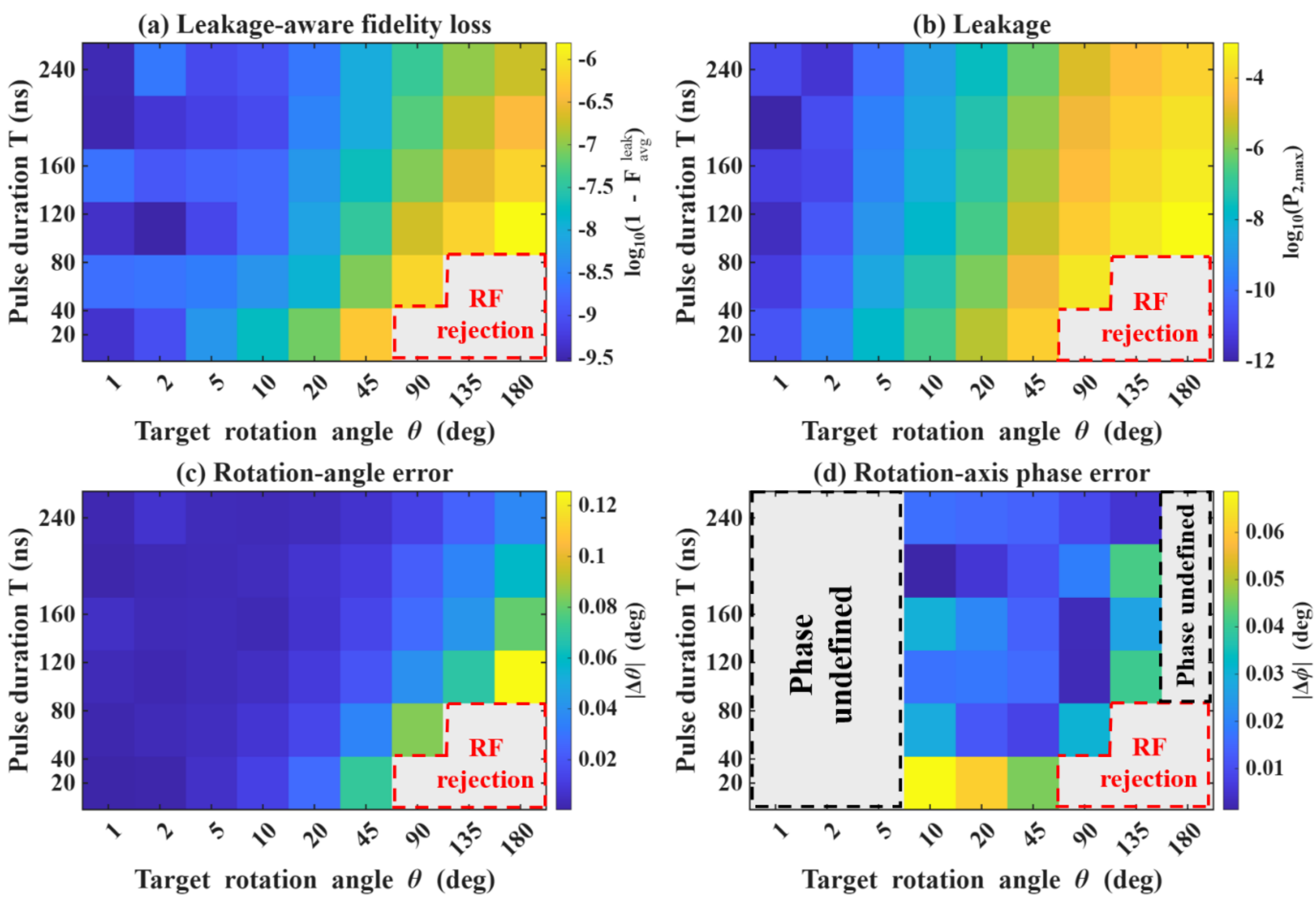

FIGURE 4. Single-qutrit closure diagnostics versus target rotation angle $\theta$ and pulse duration $T$ under the nominal direct-RF profile with $\beta = 0.5$. (a) Leakage-aware fidelity loss, $\log_{10}(1 - F_{\mathrm{avg}}^{\mathrm{leak}})$. (b) Maximum transient leakage, $\log_{10}(P_2^{\max})$. (c) Rotation-angle error $|\Delta\theta|$. (d) Rotation-axis phase error $|\Delta\phi|$. White regions outlined in red denote settings rejected by RF validation before local QuTiP evaluation. Black-dashed regions in (d) denote near-polar targets for which the rotation-axis phase is undefined and therefore not reported. Complete closure requires every applicable Table II threshold to be satisfied.

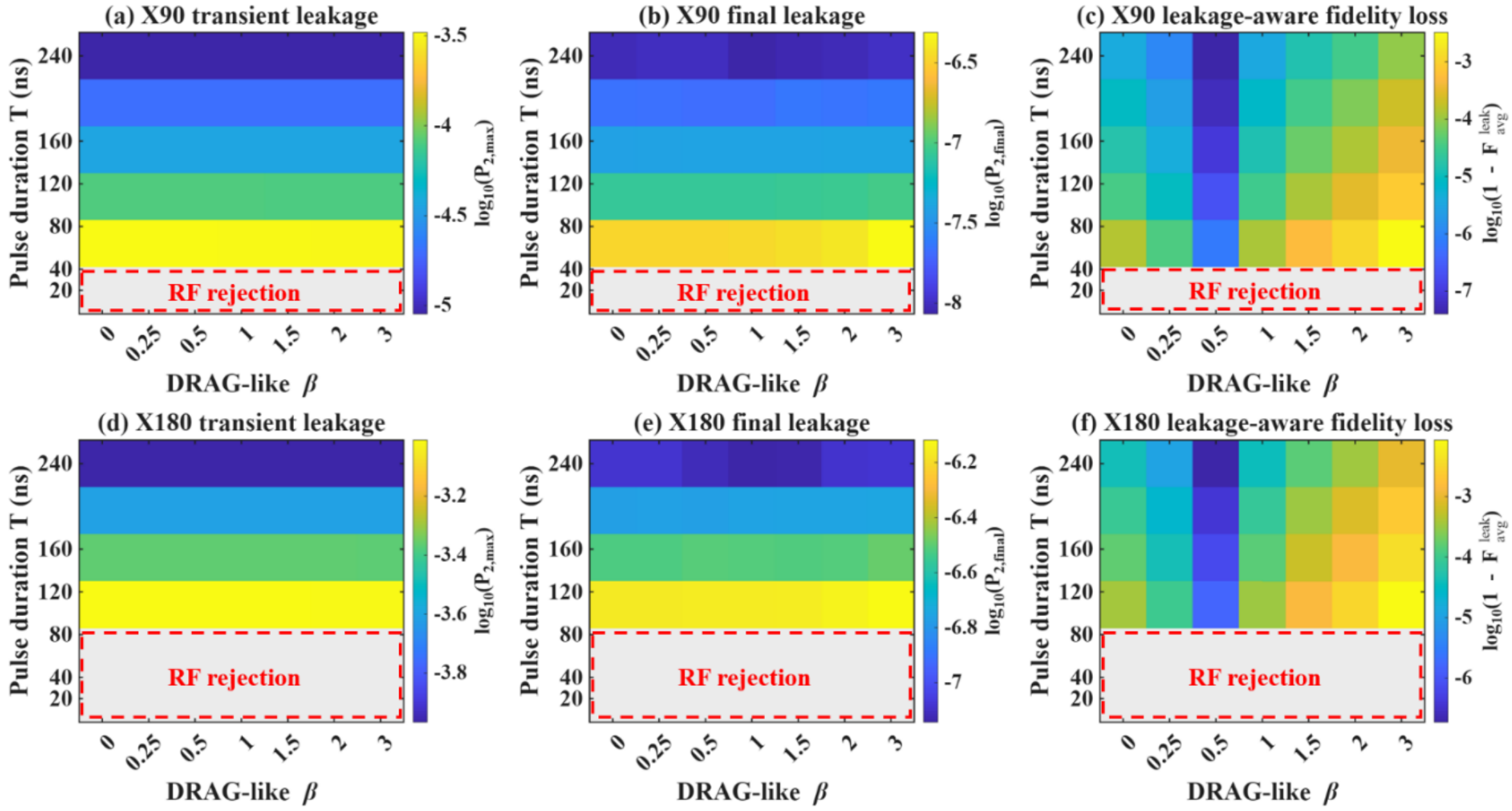

FIGURE 5. DRAG-like coefficient and pulse-duration scans for X90 and X180 targets. The minimum-useful-amplitude floor is disabled to isolate pulse-shaping physics. (a) X90 maximum transient leakage. (b) X90 final leakage. (c) X90 local leakage-aware fidelity loss. (d) X180 maximum transient leakage. (e) X180 final leakage. (f) X180 local leakage-aware fidelity loss. The β = 0.5 column minimizes the leakage-aware fidelity loss at every RF-admitted duration, whereas maximum transient leakage is governed primarily by pulse duration. Gray cells are rejected by RF validation before QuTiP evaluation.

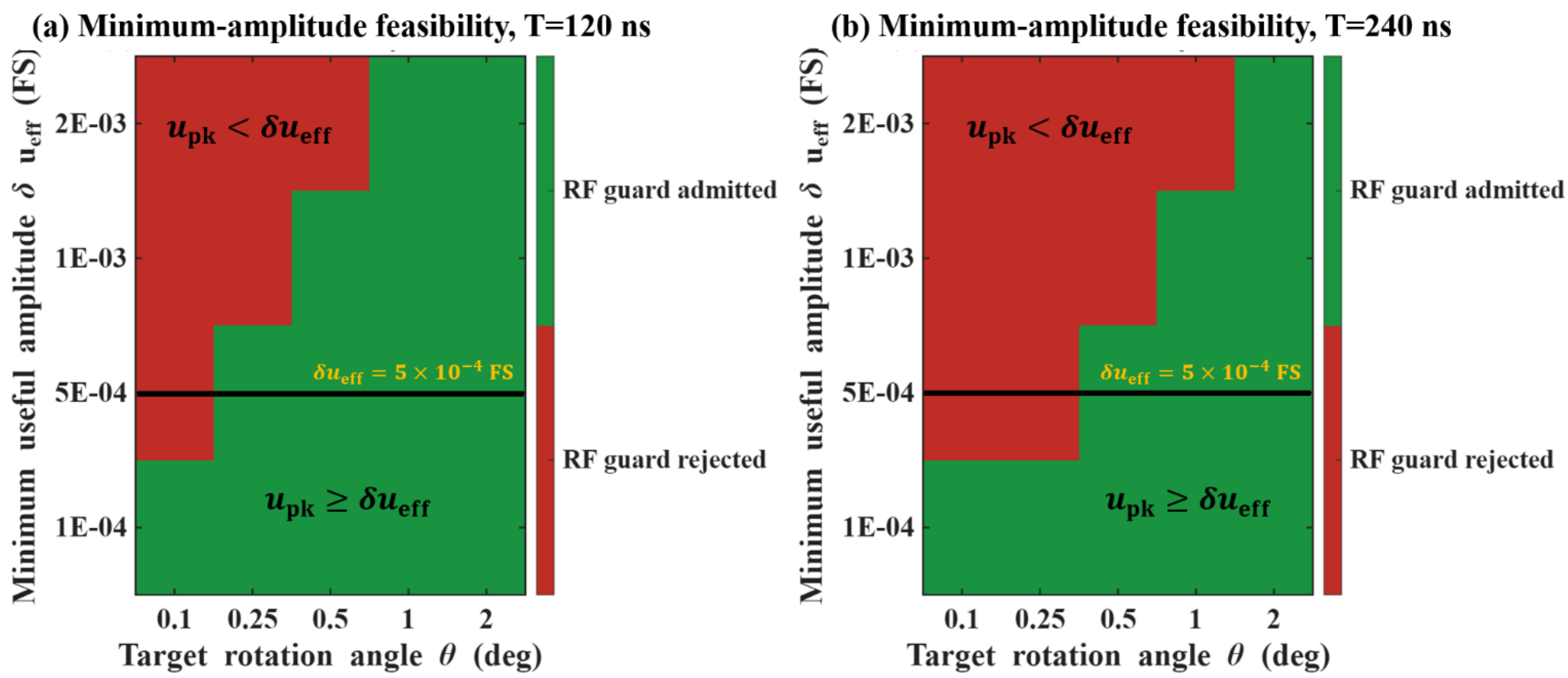

**FIGURE 6. Minimum-amplitude feasibility under the RF guard for (a) $T = 120$ ns and (b) $T = 240$ ns. The discrete sweep uses $\delta u_{\mathrm{eff}} \in \{1 \times 10^{-4}, 5 \times 10^{-4}, 1 \times 10^{-3}, 2 \times 10^{-3}\}$ FS and $\theta \in \{0.1^\circ, 0.25^\circ, 0.5^\circ, 1^\circ, 2^\circ\}$. Green cells satisfy $u_{\mathrm{pk}} \geq \delta u_{\mathrm{eff}}$ and are admitted, whereas red cells satisfy $u_{\mathrm{pk}} < \delta u_{\mathrm{eff}}$ and are rejected. The horizontal line marks the nominal Table-I value $\delta u_{\mathrm{eff}} = 5 \times 10^{-4}$ FS. The displayed staircase is defined only on this sampled grid and should not be interpreted as a continuously resolved boundary.**

### *B. Pairwise Same-Frame Coexistence Boundaries*

The scheduler requires a pairwise decision for every two requested tones before constructing the conflict graph. This pairwise study therefore evaluates the same-frame coexistence boundary using the complete RF and two-qutrit validation pipeline rather than a universal carrier-spacing rule.

Both qutrits have an anharmonicity of $\alpha = -250$ MHz. Qubit $i$ is fixed at $f_{01,i} = 5.00$ GHz, while $f_{01,j}$ is varied according to the tested sweep. Both gates are $X90$ rotations with the calibrated DRAG-like coefficient $\beta = 0.5$. The pulse duration is

$$T \in \{40,60,80,120,160,240\} \text{ ns}.$$

The minimum-useful-amplitude floor is disabled in this experiment to isolate pairwise spectral and Hamiltonian coexistence. Both tones use the same initial phase, and the RF reference frequency is placed at their midpoint. Three sweeps are performed. The carrier-spacing sweep uses

$$\Delta f_{01} \in \{20,30,40,55,75,100,150,220\} \text{ MHz}$$

with shared-line coupling $C_{ij} = C_{ji} = 1$. The leakage-transition sweep places tone $j$ around the $f_{12,i}$ transition with detunings from $-150$ to $150$ MHz, also using $C_{ij} = C_{ji} = 1$. The crosstalk sweep fixes $\Delta f_{01} = 30$ MHz and varies the symmetric coupling from zero to unity. The complete experiment contains 258 pairwise cases.

Because finite-duration tone templates are not exactly orthogonal, their post-RF complex gains are recovered by a joint least-squares projection rather than by independent matched filters. A zero-coupling restoration test gives an absolute gain error of $1.19 \times 10^{-10}$, confirming that the projection does not introduce artificial tone mixing. A separate detuning-sign regression gives

$$\frac{P_{2,\max}(\Delta = \alpha)}{P_{2,\max}(\Delta = -\alpha)} = 263.6,$$

confirming that the simulated leakage resonance occurs at the physical $f_{12}$ transition.

Figure 7(a) shows the carrier-spacing sweep. Under full shared-line coupling, only three tested points close: $\Delta f_{01} = 75$, 100, and 150 MHz at $T = 240$ ns. No tested spacing closes for $T \leq 160$ ns under all Table II criteria. At $T = 240$ ns, spacings from 20 to 55 MHz fail primarily through false addressing or rotation-axis phase error. The $\Delta f_{01} = 220$ MHz case places one carrier only 30 MHz from the other qutrit's $f_{12}$ transition and also fails. The closed corridor in the sampled grid is therefore 75-150 MHz at 240 ns. This result shows that increasing carrier separation does not guarantee closure once a carrier approaches a spectator leakage transition.

Figure 7(b) resolves the leakage-transition collision directly. At zero detuning from $f_{12,i}$, the maximum transient leakage is approximately 0.139 and is nearly independent of pulse duration. Away from resonance, the collision band narrows as the pulse duration increases. Using the $P_{2,\max} \leq 10^{-3}$ criterion, the smallest sampled two-sided passing detunings are 150, 60, 45, and 30 MHz for $T = 80$, 120, 160, and 240 ns, respectively. The corresponding largest sampled failing detunings are 120, 50, 30, and 25 MHz. These calibrated widths define the duration-dependent directed $f_{12}$ guards used by Experiments 3 and 4. They are scheduling preconditions based on transient leakage and do not replace the remaining Table II closure checks.

Figure 7(c) shows the coexistence boundary versus symmetric coupling at $\Delta f_{01} = 30$ MHz. The largest tested coupling that closes is $C_{ij} = C_{ji} = 0.002$ at 40 ns, 0.02 at 80 ns, 0.20 at 120 and 160 ns, and 0.70 at 240 ns. No coupling value closes at 60 ns because the two-tone aggregate waveform violates RF headroom. In particular, its peak is $0.977A_{\text{FS}}$, above the configured $0.891A_{\text{FS}}$ limit. The 40 ns waveform has a lower aggregate peak of $0.834A_{\text{FS}}$ because the beat-envelope maximum is differently aligned with the pulse envelope. This nonmonotonic RF behavior is deterministic and results from the zero-relative-phase crest factor. Full shared-line coupling $C_{ij} = C_{ji} = 1$ does not close at the tested 30 MHz spacing for any pulse duration.

Figure 7(d) counts the failure reasons recorded across all 258 cases. The categories are not mutually exclusive because one candidate may violate several thresholds. RF validation rejects 70 cases before qutrit evaluation. Among non-muted cases, false-addressing projection appears in 44 failures, leakage-drive projection in 78, rotation-angle error in 24, rotation-axis phase error in 111, leakage-aware fidelity loss in 46, computational-survival loss in 10, and maximum transient leakage in 47. No case violates the intended-drive-mismatch threshold. Muted RF frames are excluded from the transition-frame and QuTiP counts.

These results establish that pairwise coexistence cannot be represented by a single carrier-spacing constant. The boundary depends jointly on pulse duration, carrier spacing, leakage-transition proximity, effective coupling, aggregate RF crest factor, and the complete Table II closure criteria.

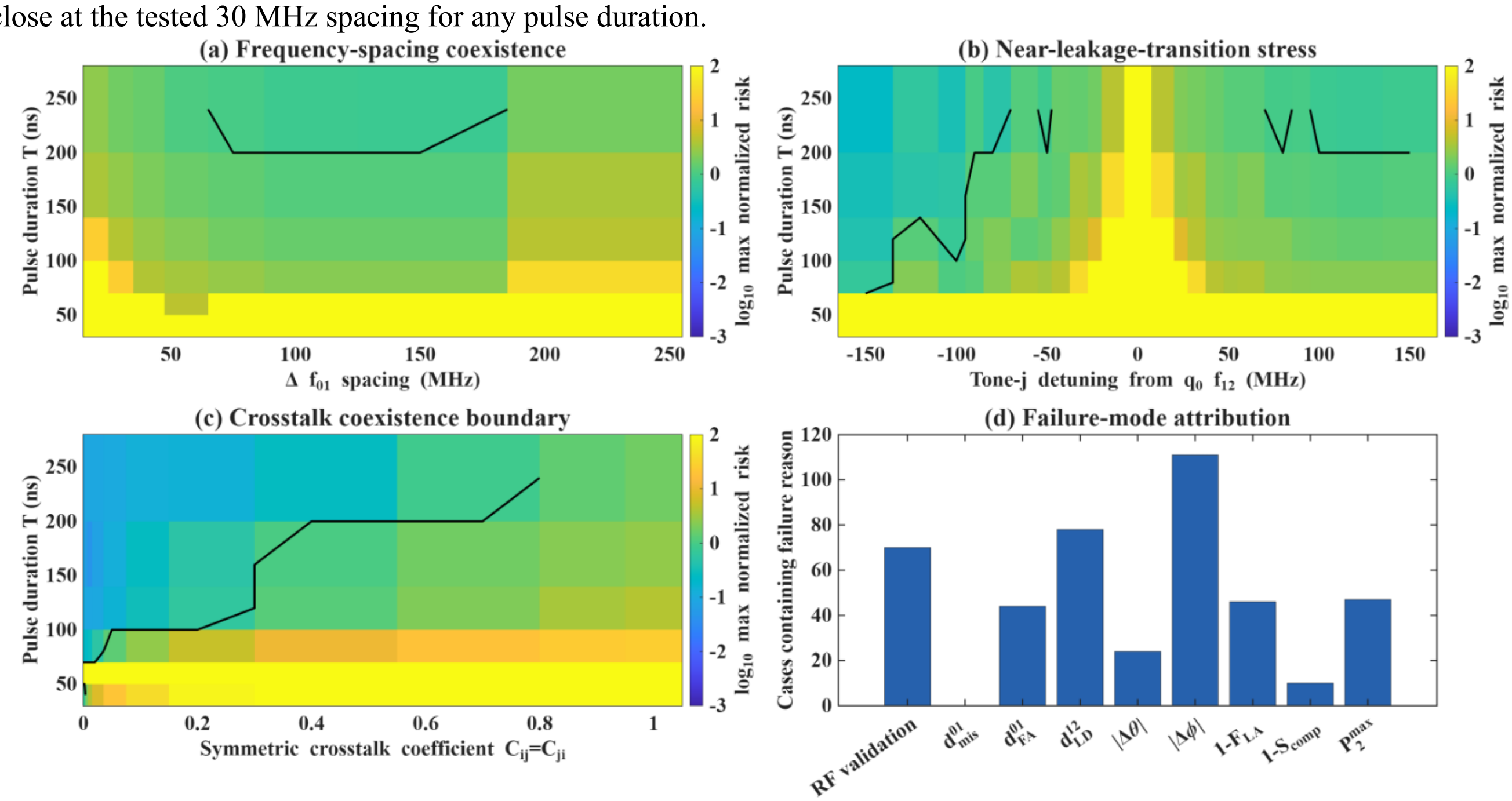

FIGURE 7. **Pairwise same-frame coexistence for two $X90$ targets with $\alpha = -250$ MHz and $\beta = 0.5$. The minimum-useful-amplitude floor is disabled, and both tones use zero relative phase. (a) Maximum normalized closure risk versus computational-transition spacing $\Delta f_{01}$ and pulse duration $T$ under full shared-line coupling $C_{ij} = C_{ji} = 1$. Small spacing produces false addressing, whereas the large-spacing boundary approaches a spectator $f_{12}$ transition. (b) Maximum normalized risk versus the detuning of tone $j$ from $f_{12,i}$. The zero-detuning leakage peak remains near $P_{2,\max} = 0.139$, while the collision band narrows with increasing $T$. (c) Maximum normalized risk versus symmetric coupling at $\Delta f_{01} = 30$ MHz. Longer pulses admit larger coupling, except where deterministic two-tone crest factor causes RF rejection. (d) Nonexclusive failure-reason counts over all 258 pairwise cases. Muted RF frames are counted only as RF-validation failures and are excluded from transition-frame and QuTiP attribution. Black contours in (a)–(c) mark the complete Table II pass/fail boundary.**

### *C. RF-Frame Capacity Under a Fixed RFSoC Budget*

Under the fixed RFSoC profile in Table I and the closure criteria in Table II, we quantify two related resource measures: the largest validated number of tones in one RF frame and the number of validated frames required to cover a complete active layer. The sweep uses a 1-GHz sub-6-GHz synthesis window from 4.75 to 5.75 GHz, active-set sizes from 2 to 16 qubits, $X90$ and $X180$ targets, and pulse durations of 80, 120, 160, and 240 ns. All pulses use the calibrated DRAG-like value $\beta = 0.5$ and a Gaussian width fraction of 0.18. The tones are assigned zero programmed relative phase; therefore, no crest-factor phase optimization is included in the reported capacities.

Four deterministic synthetic QID maps probe distinct carrier-placement regimes. The **uniform** map spaces the $f_{01}$ frequencies evenly across the 1-GHz window. The **jittered** map starts from the uniform grid and applies the deterministic perturbation $0.18\Delta f_{\text{nom}} \sin(1.7i)$, followed by sorting; it preserves broad spectral coverage while breaking regular spacing and shifting individual tones relative to leakage transitions. The **clustered** map places approximately 55% of the tones between 5.02 and 5.32 GHz and distributes the remaining tones across the full window, thereby creating a dense interior region within an otherwise broad map. The **heavy-tail** map begins from the uniform grid but inserts isolated nearest-neighbor spacings of 18, 28, and 45 MHz, representing a small number of severe spacing outliers rather than global clustering. In every map,

$$\alpha_i = -250 + 8\sin(0.9i)\ \text{MHz}, \qquad f_{12,i} = f_{01,i} + \alpha_i.$$

These maps are controlled stress instances and are not intended to reproduce a measured processor layout. Their complete QID tables are included in the reproducibility package.

All active tones are modeled as sharing one RF output, with unit off-diagonal shared-line coupling before frequency selectivity and RF-chain effects are applied. For each map, active-set size, rotation, and pulse duration, the scheduler constructs a pairwise conflict graph from the RF-frequency guard, the transition-frame projection criteria, and the duration-dependent directed $f_{12}$ guard calibrated in Section VI-B. A deterministic independent-set search proposes a candidate frame. The aggregate waveform is then encoded, propagated through the RF model, and tested against Table II. Exact local QuTiP validation is applied to the highest-risk two- to four-qutrit patch. A rejected candidate is reduced using the identified risk channel and reevaluated. Consequently, the capacities in Figs. 8(a) and 8(b) are the largest frames returned and validated by this deterministic search, not analytical pairwise-independent-set sizes or exhaustive global maxima.

For the 16-qubit $X90$ layer, Figure 8(a) shows that all four maps admit two tones at 80 ns. At 120 ns, the validated capacities for the uniform, jittered, clustered, and heavy-tail maps are 2, 3, 2, and 2 tones, respectively. They become 3, 2, 2, and 4 tones at 160 ns and 5, 5, 3, and 4 tones at 240 ns. The clustered map remains more restrictive because its dense interior region produces more simultaneous frequency and leakage-transition conflicts.

The $X180$ results in Figure 8(b) are more restrictive because the larger pulse area increases the aggregate command amplitude and leakage exposure. Every map is limited to one validated tone at 80 and 120 ns. At 160 ns, only the uniform map admits two tones, whereas the jittered, clustered, and heavy-tail maps remain at one. At 240 ns, all four maps again admit only one tone under the tested zero-relative-phase convention. Thus, pulse duration does not produce a universally monotonic capacity increase; headroom, carrier placement, leakage collisions, and the fixed phase convention act jointly.

For the complete 12-qubit $X90$ layer, a separate exact graph-coloring procedure with model-in-the-loop no-good constraints determines whether the candidate frames form a complete validated partition. No complete partition is found at 80 ns for any map because at least one singleton fails the configured closure criteria. At 120 ns, the uniform, jittered, clustered, and heavy-tail maps require $K = 5, 5, 9,$ and $5$ frames. At 160 ns, the corresponding counts are $K = 4, 5, 9,$ and $5$. At 240 ns, they decrease to $K = 3, 3, 5,$ and $3$. Figure 8(c) reports the smallest complete partition found over the tested durations: three frames for the uniform, jittered, and heavy-tail maps and five frames for the clustered map. For $B_{\text{RF}} = 1$ GHz, these 240-ns partitions correspond to $\rho_{\text{layer}} = 4$ and $2.4$ qubits/GHz/frame, respectively.

Figure 8(d) reports nonexclusive counts of the Table II failure channels across 3529 evaluated candidate frames. RF validation, leakage-drive projection, transient leakage, and rotation-axis phase error are the most frequent limiting channels. The directed $f_{12}$-guard count is zero at the candidate-validation stage because violating pairs are removed during conflict-graph construction before aggregate frames are evaluated. Since one candidate may violate multiple criteria, the bars are not mutually exclusive and do not sum to the number of rejected frames.

The frame count alone does not include the time cost of longer pulses. Figure 9 therefore reports

$$T_{\text{layer}} = KT \quad (28)$$

and the corresponding time-normalized aggregation

$$\rho_{\text{time}} = \frac{N_{\text{active}}}{B_{\text{RF}} T_{\text{layer}}}. \quad (29)$$

The 80-ns settings are omitted because they do not yield complete validated partitions. At 120 ns, the uniform, jittered, and heavy-tail maps have $T_{\text{layer}} = 600$ ns and $\rho_{\text{time}} = 20$ qubits/GHz/μs, while the clustered map gives 1080 ns and 11.11 qubits/GHz/μs. At 160 ns, the corresponding time-normalized values are 18.75, 15, 8.33, and 15 qubits/GHz/μs. At 240 ns, they are 16.67, 16.67, 10, and 16.67 qubits/GHz/μs. Hence, 240 ns minimizes the number of frames, whereas 120 ns maximizes the time-normalized aggregation for all four tested maps.

These results establish simulation-based, budget-dependent capacities for the specified synthetic QID maps. They are not measured channel capacities or hardware wiring-reduction factors. The QuTiP stage validates local two- to four-qutrit risk patches rather than a fully coupled 16-qutrit Hamiltonian.

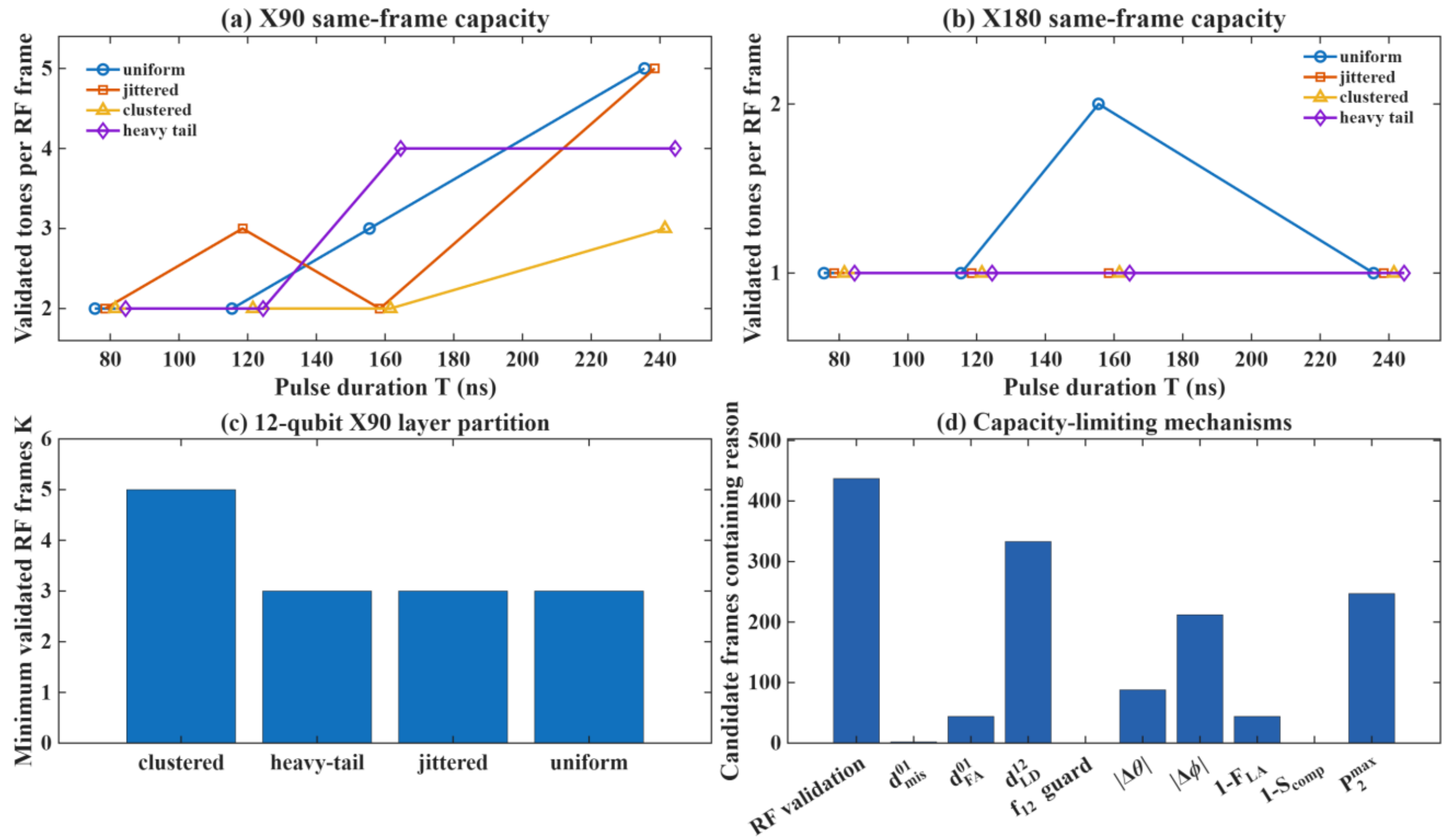

**FIGURE 8.** RF-frame capacity under the fixed Table I RFSoC profile and Table II closure criteria. The deterministic synthetic maps cover uniform spacing, deterministic spacing perturbations, a dense interior cluster, and isolated heavy-tail spacing outliers within 4.75–5.75 GHz. (a) Largest validated X90 frame returned for the 16-qubit maps. (b) Corresponding X180 results. (c) Minimum number $K$ of frames in a complete validated partition of the 12-qubit X90 layer over the tested pulse durations. (d) Nonexclusive counts of candidate frames containing each failure reason. Directed-$f_{12}$-guard violations are absent from panel (d) because such pairs are removed during conflict-graph construction. Candidate frames are propagated through the aggregate RF model and validated using exact local two- to four-qutrit risk patches; the plotted capacities are not fully coupled 16-qutrit simulations.

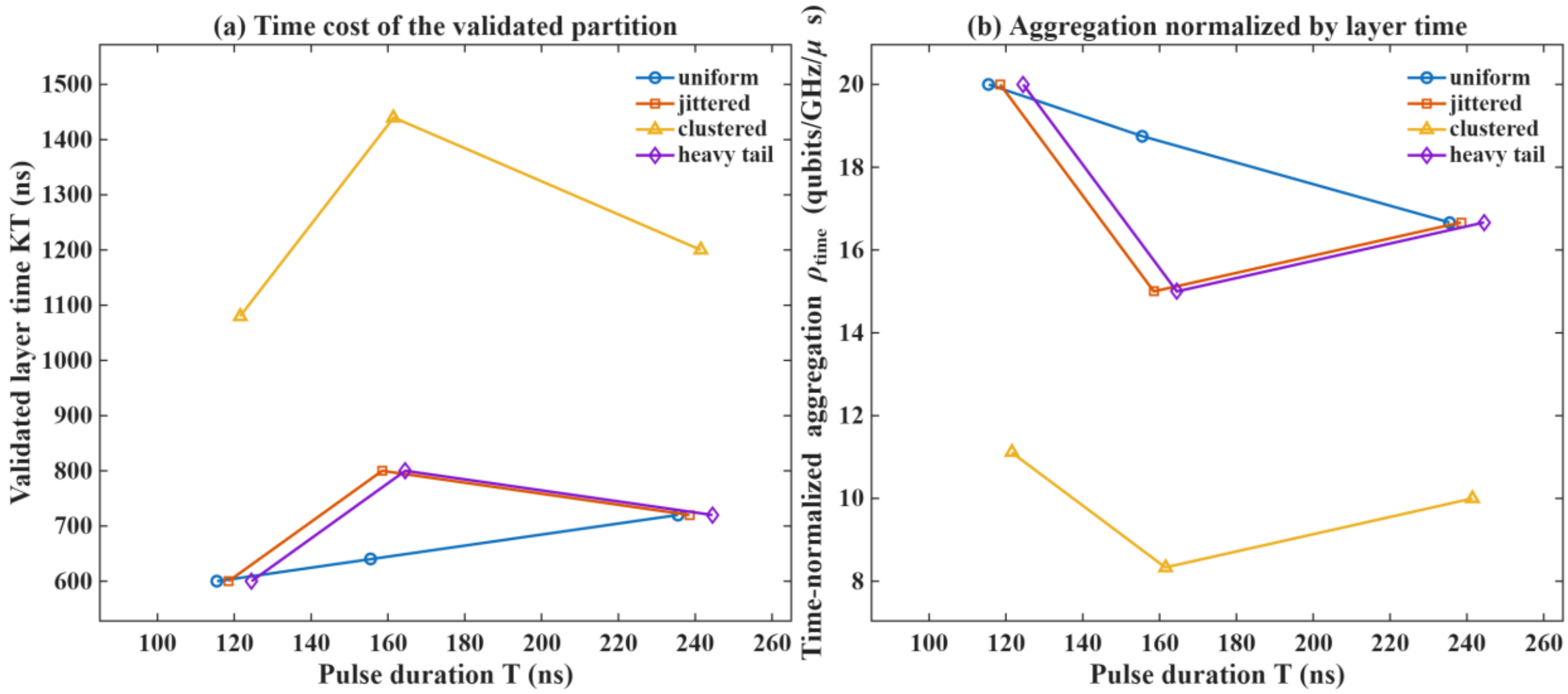

**FIGURE 9.** Time cost of the complete validated 12-qubit X90 partitions. (a) Validated layer time $T_{layer} = KT$. (b) Time-normalized aggregation $\rho_{time} = N_{active}/(B_{RF}T_{layer})$. Settings without a complete validated partition are omitted; no map closes at 80 ns. Although 240 ns reduces the required frame count, 120 ns provides the largest time-normalized aggregation for all four tested maps. Small horizontal marker offsets are included only for visual separation.

### *D. Algorithm-Derived Microwave Workloads: Bernstein–Vazirani and QAOA*

The preceding subsections used controlled synthetic layers to isolate same-frame coexistence and RF-frame capacity. We next apply the same scheduling, RF propagation, and qutrit-level closure workflow to microwave layers extracted from Qiskit-generated Bernstein–Vazirani (BV) and QAOA circuits [30], [33], [34]. Figure 10 separates two distinct workload classes. In BV, every data qubit receives the same fixed physical $R_y(-\pi/2)$ rotation in each Hadamard-derived layer. In QAOA, only the mixer layer is assigned to the shared RF output, and its common $R_x(\theta_M)$ rotation angle is varied. BV oracle CNOTs, ancilla preparation, QAOA state preparation and cost-layer operations, and readout remain outside the shared-RF workloads evaluated here. The results therefore characterize algorithm-derived single-qubit microwave layers rather than end-to-end algorithm execution.

For BV, each Hadamard operation is represented, up to a global phase, as

$$H \doteq R_z(\pi)R_y(-\pi/2). \quad (30)$$

With the rightmost operation applied first, Eq. (30) is implemented by a fixed physical $R_y(-\pi/2)$, or $-Y90$, microwave rotation followed by a zero-duration virtual $R_z(\pi)$ frame update. Only the physical $-Y90$ layer consumes RF-frame time. The virtual update is retained in the phase bookkeeping, and the terminal update before computational-basis measurement is absorbed into the readout frame.

For QAOA, the shared-RF workload is the mixer unitary

$$U_M(\beta_{\mathrm{QAOA}}) = \exp\left(-i\beta_{\mathrm{QAOA}}\sum_j X_j\right) = \otimes_j R_x\left(2\beta_{\mathrm{QAOA}}\right), \quad (31)$$

thus, each mixer layer becomes an equal-angle $R_x(\theta_M)$ workload, where $\theta_M = 2\beta_{\mathrm{QAOA}}$. Unlike the fixed $-Y90$ BV layers, $\theta_M$ is swept to expose the dependence of RF-frame aggregation on the requested physical rotation angle. State preparation and the cost unitary $U_C(\gamma)$ are outside the shared-RF workload.

Each extracted microwave layer must be covered by disjoint RF frames that individually satisfy the Table I RF conditions and all applicable Table II closure criteria. Candidate frames are generated from the duration-dependent conflict graph using the scheduler described in Sec. IV [35] and are then propagated through the aggregate RF model and checked with local qutrit simulations. The reported $K$ is the number of validated frames in a complete layer partition. For selected settings, all candidate subsets up to the headroom-feasible frame size are enumerated, and a minimum disjoint cover is solved over the validated frame library. These exact-cover audits distinguish a physically validated minimum from a scheduler-produced feasible partition.

Fixed-angle BV workload. Figure 11(a) shows the audited 12-qubit BV schedule at $T = 240$ ns. The exact-cover audit gives $K = 3$ four-tone RF frames for each fixed $-Y90$ layer. The two Hadamard-derived layers therefore require six shared-RF bursts, arranged on the two sides of the oracle. The three distinct aggregate peaks are approximately $0.50A_{\mathrm{FS}}$, $0.50A_{\mathrm{FS}}$, and $0.47A_{\mathrm{FS}}$, all below the configured headroom limit $b_{\mathrm{BO}}A_{\mathrm{FS}} = 0.891A_{\mathrm{FS}}$.

Figure 12 reports the fixed-angle BV scaling results. At $T = 120$ ns, all tested workloads from 4 to 16 active qubits admit complete partitions, with $K$ increasing from 2 to 7. At $T = 240$ ns, complete partitions are found for $N = 4,6,10,12,$ and 14 with $K = 2,2,3,3,$ and 4, respectively; the independently generated 8- and 16-qubit maps do not admit complete validated covers. For the 12-qubit layer, increasing the pulse duration from 120 to 240 ns reduces $K$ from 5 to 3, but increases the validated microwave-layer time $KT$ from 600 to 720 ns. Longer pulses therefore improve tones-per-frame aggregation without necessarily minimizing layer time. Dashed curves in Figure 12(a) show the pairwise-capacity lower bound, and the dagger identifies an exact-cover audit.

Variable-angle QAOA mixer workload. Figure 11(b) gives a concrete 12-qubit $X120$ mixer layer at $T = 240$ ns. Its exact validated partition contains $K = 5$ frames with $2,2,2,3,$ and 3 tones. Their aggregate peaks are $0.35A_{\mathrm{FS}}$, $0.34A_{\mathrm{FS}}$, $0.34A_{\mathrm{FS}}$, $0.43A_{\mathrm{FS}}$, and $0.44A_{\mathrm{FS}}$, all below the headroom limit. Across the five validated frames, the worst-case diagnostics are $P_2^{\max} = 3.81\times10^{-4}$, $|\Delta\theta| = 0.12^\circ$, and $|\Delta\phi| = 0.91^\circ$, where each quantity is maximized independently over the five frames.

Figure 13 reports the variable-angle QAOA results for a 12-qubit mixer layer. At $T = 120$ ns, $K = 4$ for $15^\circ$–$60^\circ$, followed by $K = 5,6,7,$ and 12 at $90^\circ, 120^\circ, 150^\circ,$ and $180^\circ$. At $T = 240$ ns, $K = 3$ for $30^\circ$ – $60^\circ$ under scheduler validation, while the exact-cover-audited $90^\circ$, $120^\circ$, $150^\circ$, and $180^\circ$ cases require $K = 3$, 5, 6, and 5, respectively. The corresponding layer aggregation $\rho_{\mathrm{layer}}$ and time-normalized aggregation $\rho_{\mathrm{time}}$ show that increasing the physical mixer angle generally reduces aggregation through increased pulse area, headroom demand, and leakage exposure. The apparent improvement from $150^\circ$ to $180^\circ$ is not interpreted as a general physical trend: at the polar $180^\circ$ target, the rotation-axis phase is undefined and its phase constraint is inactive.

Figure 14 compares selected worst-frame diagnostics for the validated fixed-angle BV and variable-angle QAOA partitions at $T = 240$ ns. For the exact BV partition, the maximum rotation-angle error is $0.088^\circ$, the maximum phase error is $0.800^\circ$, and the maximum transient leakage is $1.74\times10^{-4}$. Its complete normalized risk is 0.959 and is set by the directed leakage-drive diagnostic. For the $180^\circ$ QAOA partition, the representative angle error is $0.929^\circ$ and the maximum transient leakage is $6.24\times10^{-4}$. All 100 robustness seeds close, corresponding to a Wilson 95% confidence interval of $[0.963, 1.000]$; the close agreement between the deterministic value $0.928^\circ$ and the stochastic median $0.929^\circ$ indicates that the remaining angle-error margin is dominated by deterministic command-grid and RF-chain effects.

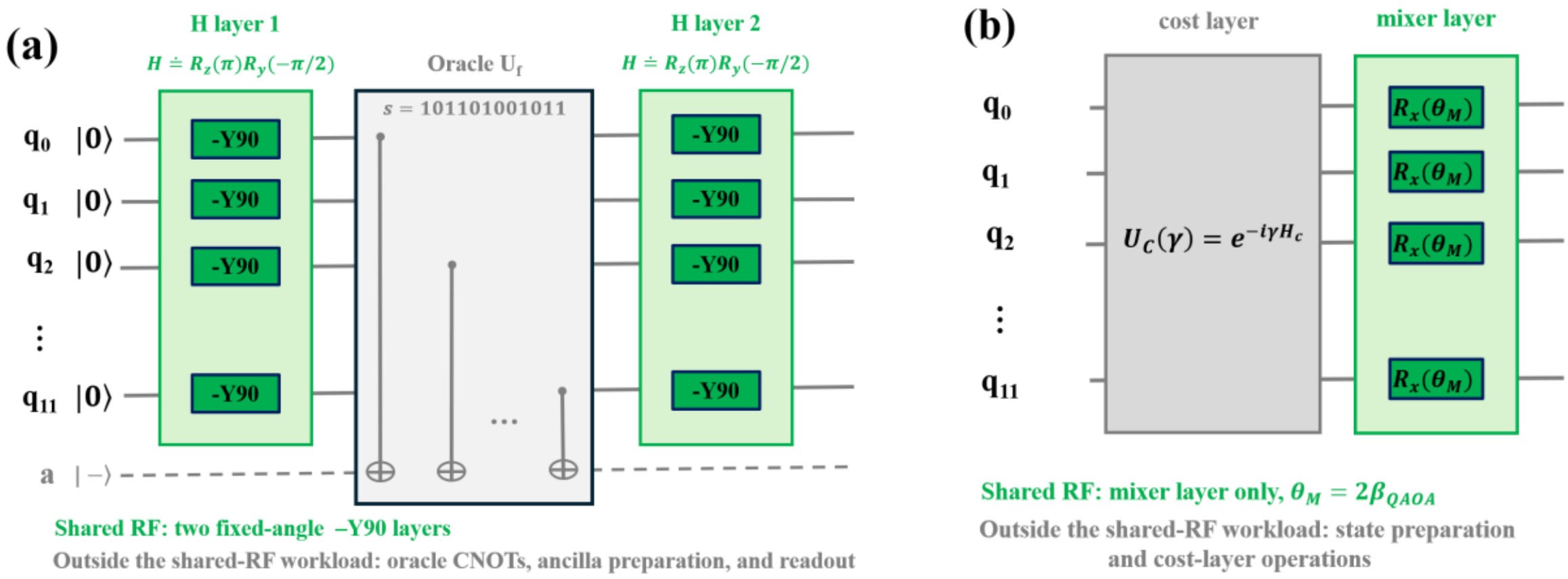

**FIGURE 10.** Algorithm-derived shared-RF workloads. (a) BV uses two fixed physical $R_y(-\pi/2)$ microwave layers; oracle CNOTs, ancilla preparation, and readout remain outside the evaluated shared-RF workload. (b) QAOA assigns only the variable-angle $R_x(\theta_M)$ mixer layer to the shared RF output, with $\theta_M = 2\beta_{\mathrm{QAOA}}$; state preparation and the cost layer remain outside the evaluated workload.

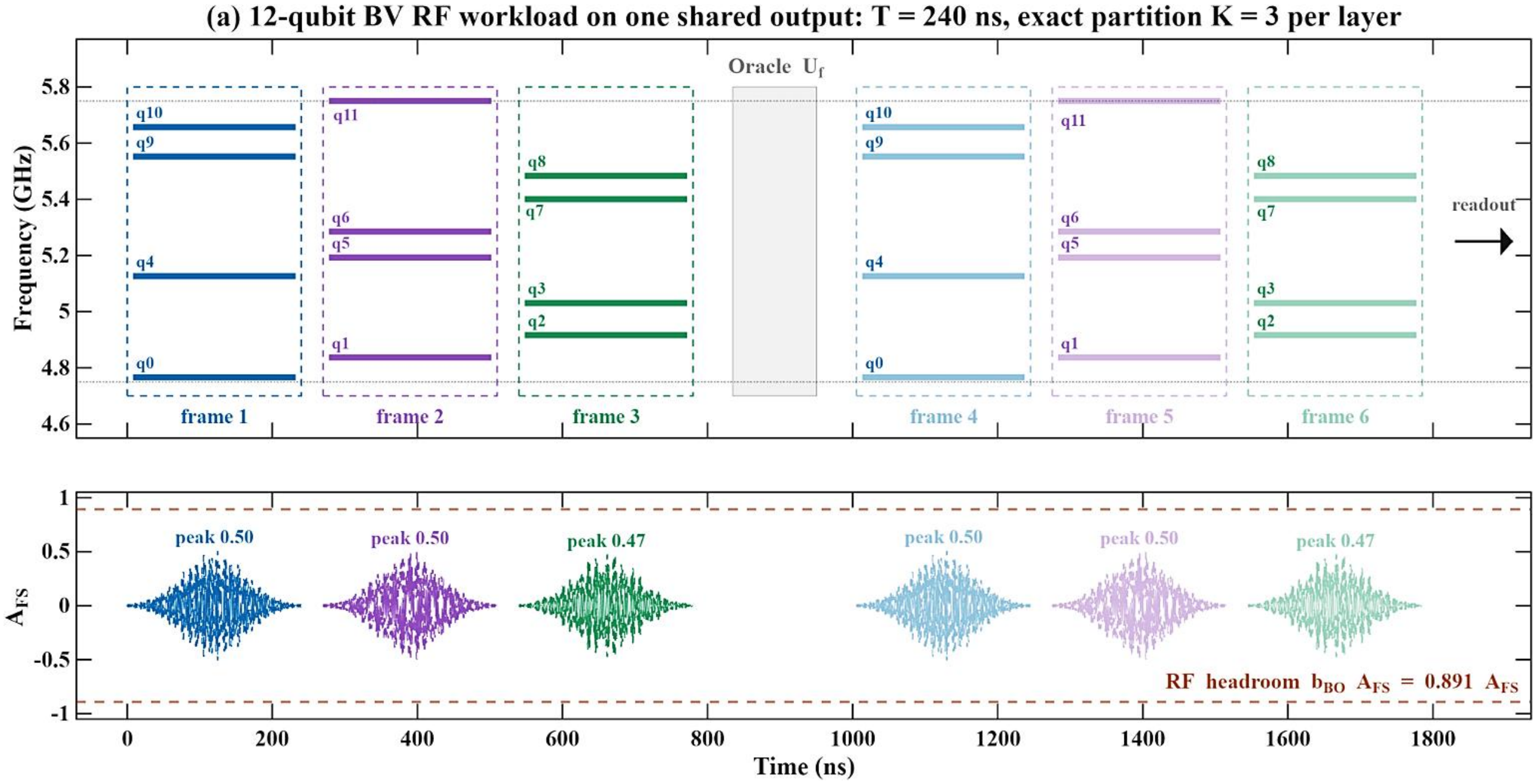

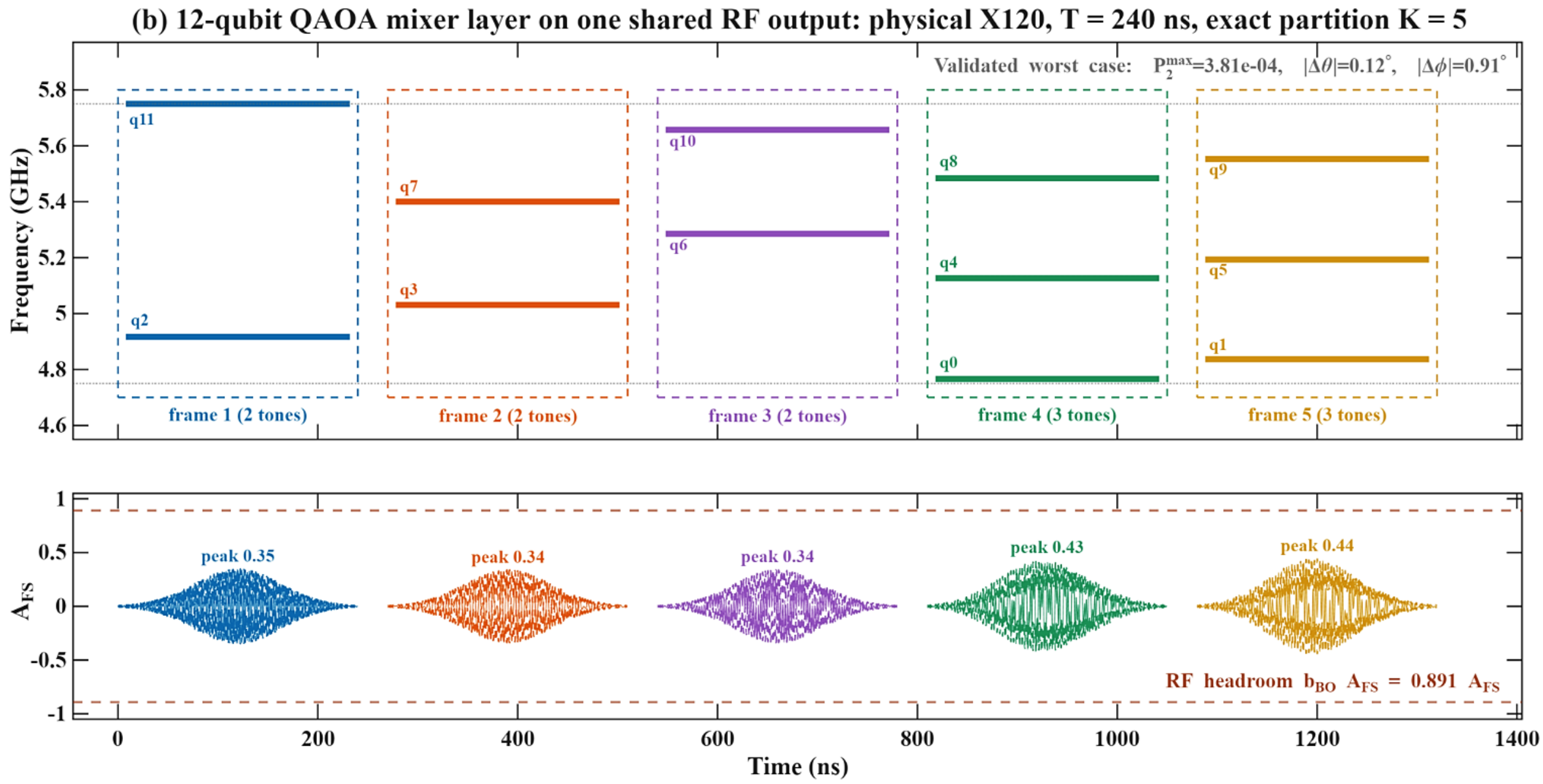

**FIGURE 11.** Concrete validated RF-frame partitions at $T = 240$ ns. (a) The 12-qubit BV workload uses three four-tone frames for each fixed $-Y90°$ layer. (b) The 12-qubit $X120$ QAOA mixer layer uses five frames containing $2, 2, 2, 3,$ and 3 tones. The lower panels show the aggregate command magnitudes; every frame remains below $b_{BO}A_{FS} = 0.891A_{FS}$.

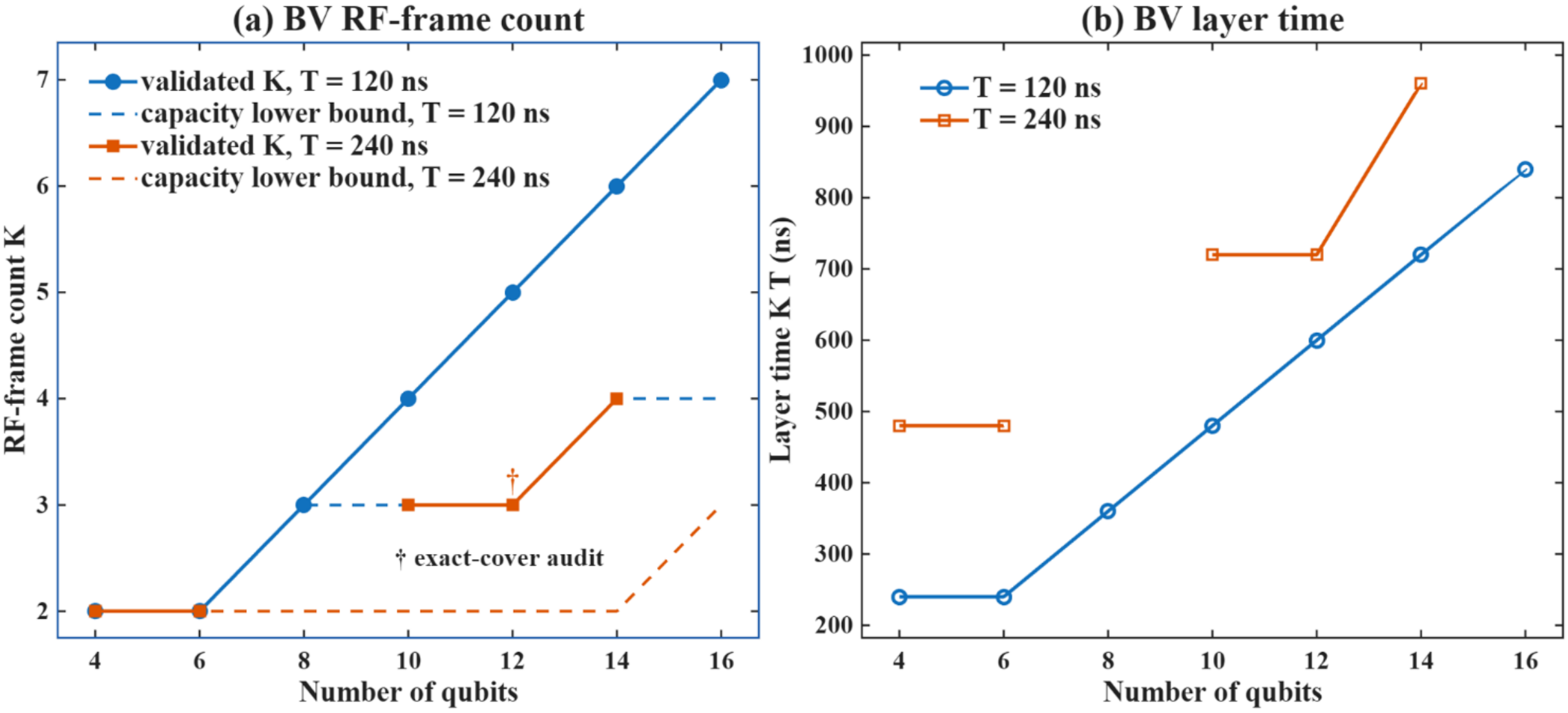

**FIGURE 12.** Fixed-angle BV partition results. (a) Validated RF-frame count $K$ versus the number of active qubits for $T = 120$ and 240 ns. Dashed curves show pairwise-capacity lower bounds, and the dagger identifies an exact-cover audit. Missing points denote settings without a complete validated partition. (b) Corresponding validated microwave-layer time $KT$.

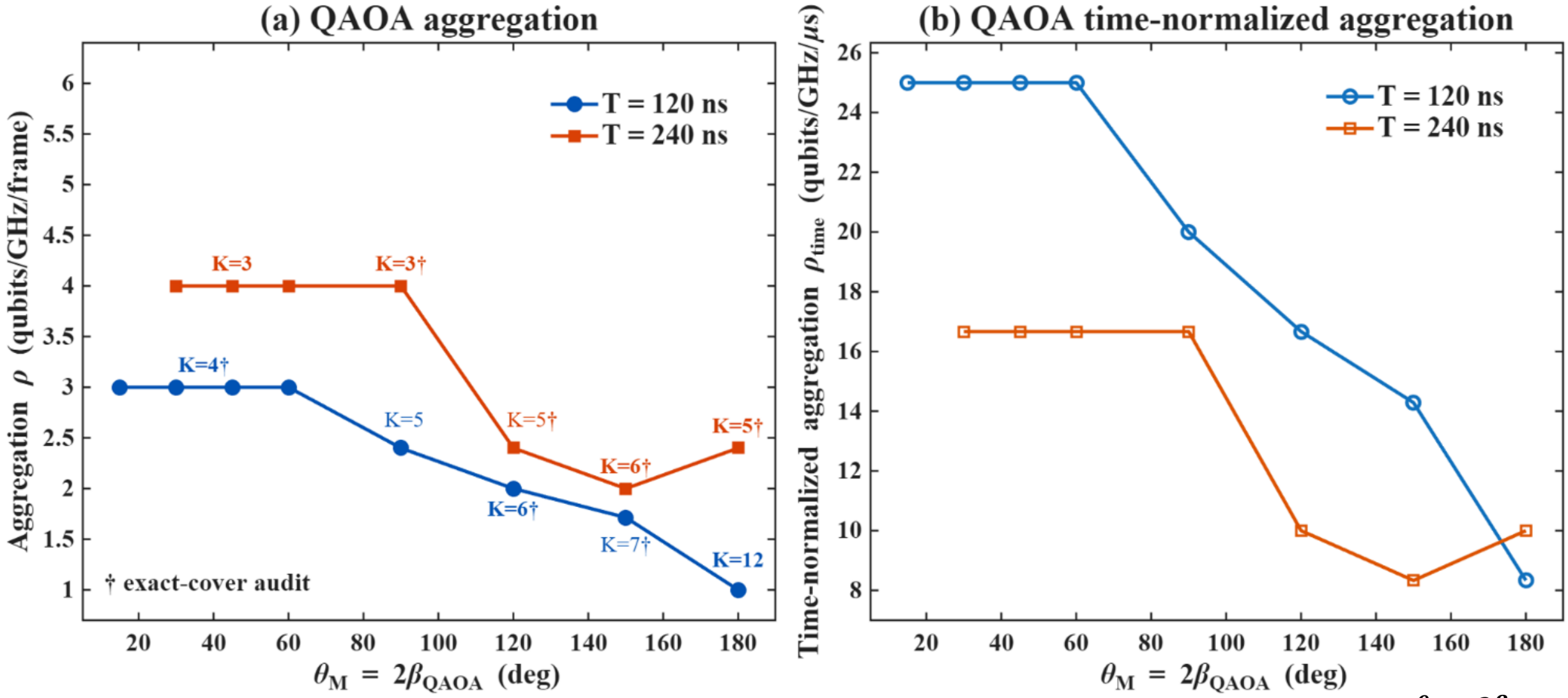

**FIGURE 13.** **Variable-angle QAOA mixer-layer results for 12 qubits. (a) Layer aggregation $\rho_{\text{layer}}$ versus the physical mixer angle $\theta_M = 2\beta_{\text{QAOA}}$; labels report the validated frame count $K$, and daggers identify exact-cover audits. (b) Corresponding time-normalized aggregation $\rho_{\text{time}}$. At the polar $180^\circ$ target, the rotation-axis phase is undefined and its phase constraint is inactive.**

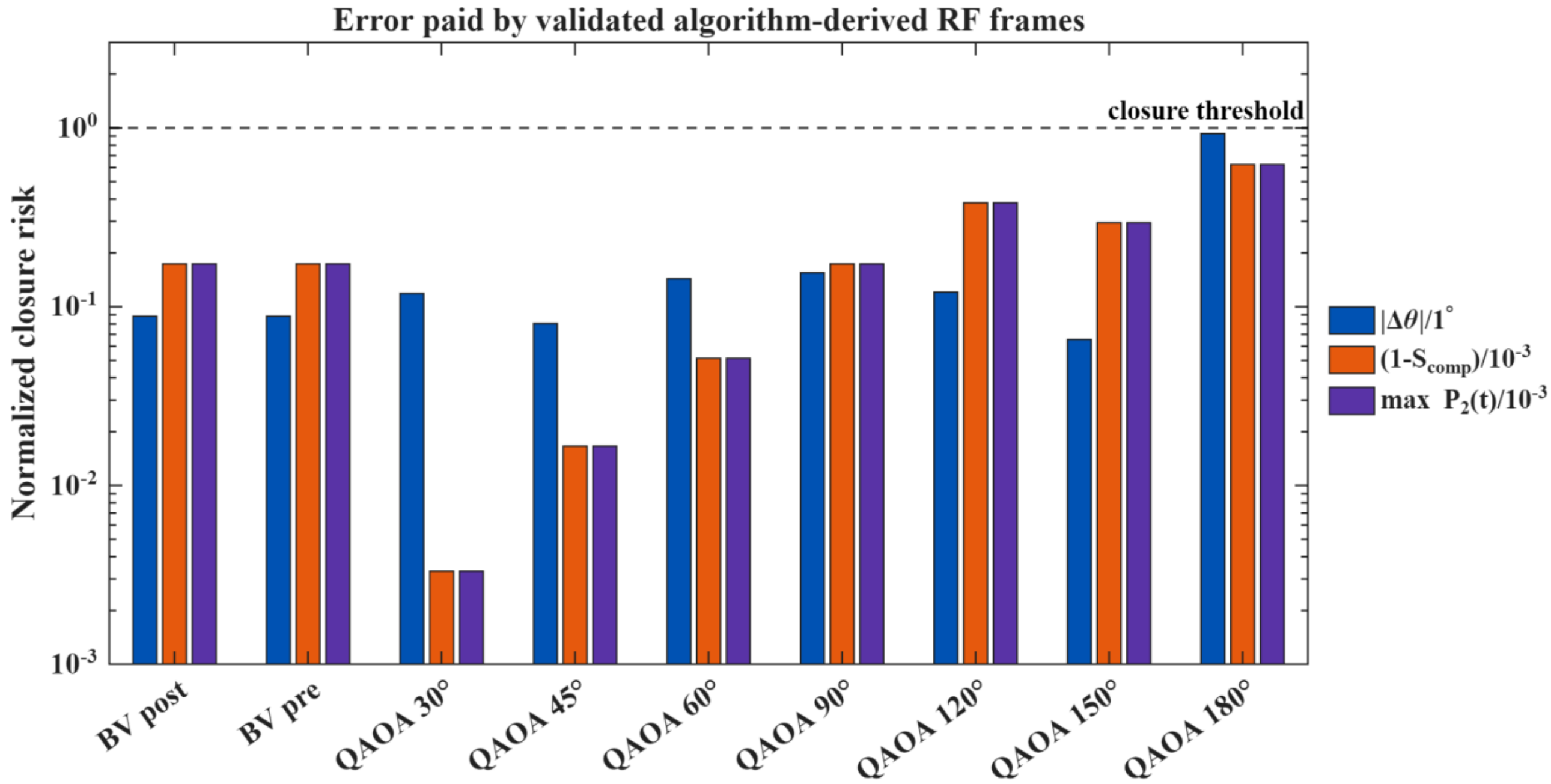

**FIGURE 14.** **Selected worst-frame diagnostics for the validated fixed-angle BV and variable-angle QAOA partitions at $T = 240$ ns. Rotation-angle error, computational-survival loss, and maximum transient $|2\rangle$ population are normalized by their Table II thresholds; unity denotes the corresponding closure limit. Each bar is the maximum over all frames in the validated partition. The complete closure decision also includes RF validation, phase error where applicable, and the remaining Table II diagnostic channels, which are not all displayed here.**

### *E. Engineering Implications and Outlook*

**Control hardware**. Aggregate-waveform headroom is the principal hard constraint within the RF-side validation stage, whereas complete frame closure can additionally be limited by directed leakage drive, rotation-axis phase error, and local qutrit diagnostics. The 14-bit DAC quantization and 48-bit RFDC NCO granularity do not become binding in the tested configurations; this observation does not apply to the effective command-resolution floor $\boldsymbol{\delta u_{\text{eff}}}$, which remains relevant for small rotations. All capacity results use the conservative zero-relative-phase convention. Per-tone phase scheduling may reduce the crest factor and recover RF headroom, but it also changes coherent false-addressing and leakage contributions. Any phase-optimized frame must therefore be re-evaluated using the same Table-I model and Table-II closure criteria. Among the tested pulse durations, the time-normalized results favor the shortest duration for which a complete validated partition exists.

**Frequency allocation**. The QID frequency map strongly affects shared-frame capacity. For a pair of qubits $\boldsymbol{i}$ and $\boldsymbol{j}$, the directed separations $|f_{01,i} - f_{12,j}|$ and $|f_{01,j} - f_{12,i}|$ must remain outside the duration-dependent leakage guard $g_{\mathrm{LD}}(T)$. In the 12-qubit $X90$ study, the clustered map requires nine rather than five validated frames at 120 ns and five rather than three frames at 240 ns, corresponding to a frame-count increase of approximately $1.7$–$1.8\times$. The best tested maps require $K = 3$ at 240 ns, whereas the clustered map requires $K = 5$. These results motivate using validated frame count as a co-objective in chip-frequency planning, together with entangling-gate collision constraints. Pairwise spacing and headroom checks alone are insufficient because coherent multitone leakage can invalidate an otherwise admissible partition; any proposed frequency allocation must therefore pass aggregate RF and local-qutrit validation before its capacity is reported.

**Scope**. The reported capacities are simulation-based, control-induced, and decoherence-free results under the Table-I RF model and Table-II closure budgets. Incorporating open-system $\boldsymbol{T_1/T_2}$ dynamics into local validation, anchoring the RF profile to measured hardware, co-designing entangling-gate layers, and experimentally calibrating the complete signal path are natural next steps.

## VII. CONCLUSION

This work formulates shared direct-RF qubit control as a validated RF-frame partitioning problem. QID records provide qubit-specific transition, pulse, and drive-calibration information; the separate RF profile specifies the configured hardware and signal-path constraints, and $\mathbf{C}$ represents the effective crosstalk coupling. A bounded exact-coloring scheduler based on Ref. [35] proposes minimum-frame partitions from the pairwise conflict graph. Each candidate frame is then encoded without pulse-area renormalization, propagated through the Table-I behavioral RF chain, and evaluated against the Table-II closure criteria and local QuTiP qutrit-patch diagnostics. Rejected nonlinear frame combinations are returned as no-good constraints for recoloring. The resulting output is therefore a validated frame count and an attributed failure mechanism, rather than an unrestricted simultaneous-control claim.

The numerical studies establish this workflow from single-qubit closure to algorithm-derived microwave layers. Single-qutrit sweeps identify the pulse-duration dependence of local closure and a DRAG-like coherent-error minimum near $\beta = 0.5$. Pairwise simulations calibrate directed leakage-transition guards of 150, 60, 45, and 30 MHz for pulse durations of 80, 120, 160, and 240 ns, respectively. For 12-qubit $X90$ layers over a 1-GHz control band, the uniform, jittered, and heavy-tail maps require five validated frames at 120 ns and three at 240 ns, whereas the clustered map requires nine and five frames, respectively. Although the longer pulse admits more tones per frame, the time-normalized aggregation metric is highest at 120 ns, demonstrating that frame capacity alone does not determine control-layer cost.

The same compiler boundary is retained for Qiskit-derived workloads. A 12-qubit Bernstein–Vazirani Hadamard-derived microwave layer, implemented as physical $-Y90$ rotations followed by virtual-$Z$ updates, closes at 240 ns in three validated four-tone frames. This partition gives $\rho_{\mathrm{layer}} = 4$ qubits/GHz/frame, with a maximum rotation-angle error of $0.088^\circ$, a maximum applicable rotation-axis phase error of $0.800^\circ$, and a maximum transient leakage population of $1.74 \times 10^{-4}$. QAOA mixer layers exhibit the expected dependence on physical rotation angle and pulse duration. The audited $180^\circ$, 240-ns partition satisfies all applicable criteria for 100 of 100 stochastic realizations, with a Wilson 95% lower confidence bound of 0.963.

These results are simulation-based, control-induced, and decoherence-free capacity estimates under the stated QID maps, Table-I behavioral RF model, and Table-II closure budgets. They do not constitute measured ZCU216 performance, measured gate fidelity, a hardware wiring-reduction factor, or validation of entangling-gate and readout resources. The principal contribution is a reproducible method for determining how many qubit-control tones can share one RF frame, when additional frames are required, and which QID or RF constraint limits closure. Measured-path anchoring, open-system $T_1/T_2$ dynamics, validated crest-factor-aware phase scheduling, and joint treatment of entangling-gate layers are the next steps toward hardware deployment.

## ACKNOWLEDGMENT

This material is based upon work supported by the U.S. Department of Energy, Office of Science, Office of Nuclear Physics, under Contract No. 89243126CSC000213.